\documentclass[reprint,amsmath,amssymb,aps]{revtex4-1}

\usepackage{graphicx}
\usepackage{dcolumn}
\usepackage{bm}
\usepackage{mathtools}
\usepackage{tikz}
\usepackage{stackengine}
    
\DeclarePairedDelimiter\bra{\langle}{\rvert}
\DeclarePairedDelimiter\ket{\lvert}{\rangle}

\newrobustcmd*{\mysquare}[1]{\tikz{\filldraw[draw=#1,fill=#1] (0,0) rectangle (0.15cm,0.15cm);}}
\newrobustcmd*{\mycircle}[1]{\tikz{\filldraw[draw=#1,fill=#1] (0,0) circle [radius=0.08cm];}}
\newrobustcmd*{\mytriangle}[1]{\tikz{\filldraw[draw=#1,fill=#1] (0,0) -- (0.2cm,0) -- (0.1cm,0.2cm);}}

\begin{document}

\preprint{APS/123-QED}

\title{Chemical Bonding in Metallic Glasses from Machine Learning\\and Crystal Orbital Hamilton Population}

\author{Ary R. Ferreira}
\email{ary.ferreira@df.ufscar.br}
\affiliation{Department of Physics, Universidade Federal de S\~{a}o Carlos (UFSCar), 13565--905, S\~{a}o Carlos, SP, Brazil}

\date{\today}

\begin{abstract}
The chemistry (composition and bonding information) of metallic glasses (MGs) 
is at least as important as structural topology for understanding their 
properties and production/processing peculiarities. This article reports a 
machine learning (ML)-based approach that brings an unprecedented ``big picture'' 
view of chemical bond strengths in MGs of a prototypical alloy system. The 
connection between electronic structure and chemical bonding is given by crystal 
orbital Hamilton population (COHP) analysis; within the framework of density 
functional theory (DFT). The stated comprehensive overview is made possible through 
a combination of: efficient quantitative estimate of bond strengths supplied by 
COHP analysis; representative statistics regarding structure in terms of atomic 
configurations achieved with classical molecular dynamics simulations; and the 
smooth overlap of atomic positions (SOAP) descriptor. The study is supplemented by 
an application of that ML model under the scope of mechanical loading; in which 
the resulting overview of chemical bond strengths revealed a chemical/structural 
heterogeneity that is in line with the tendency to \textit{bond exchange} 
verified for atomic local environments. The encouraging results pave the way 
towards alternative approaches applicable in plenty of other contexts in which 
atom categorization (from the perspective of chemical bonds) plays a key role.
\end{abstract}

\maketitle

\section{Introduction}

Since early reports of glassy alloys, over almost 60 years 
ago~\cite{KlementWillens1960}, their importance within the broad scope 
of technological developments of metallic materials has grown noticeably; 
albeit such pertinence has apparently reached a threshold in last 
decade~\cite{MilanezFaria2017}. Metallic glasses (MGs) are amorphous 
alloys that exhibit a glass transition and are notorious for their 
extreme hardness and strength; thus they became obvious candidates for 
structural applications, while other relevant usage proposals have 
already been put forward in domains like biomedicine, nanotechnology, 
and energy~\cite{Chen2011,JiaoLiu2017}. Yet, 
at the present time, research and development (R\&D) activities related 
to this class of advanced materials are still facing challenging 
issues such as critical casting thickness, or brittleness of some nominal 
compositions found to be good glass formers. Nevertheless, recent advances 
in basic and applied research have provided innovative strategies for 
designing new MGs; with the focus on enhanced mechanical, chemical, 
and magnetic properties. For example, 3D printing (or additive manufacturing) 
is emerging as a promising alternative for the fabrication of Fe-based 
MGs parts; to be employed as magnetic shielding or transformer (electrical 
device) core laminations~\cite{MahboobaThorsson2018,LiZhang2018}. Another 
prospect, relevant for energy applications, is the synthesis of MGs with 
large specific surface areas; found to be efficient for ultrafast 
hydrogen uptake~\cite{JiaoLiu2017}.

Over the last decades, technological developments in characterization 
techniques were crucial to promote insights into the structure of 
multicomponent MGs within the full range of length scales. Down 
to the atomistic scale, the lack of long-range crystalline order (often 
accompanied by nontrivial chemistry) makes it a hard task to uncover 
mechanisms underlying transformations occurring in two essential contexts: 
(i) one is the production process, what requires mastery of the key 
factors (or formation mechanisms) that determine glass-forming ability 
(GFA); (ii) the other scenario is application and concerning their most 
prominent usability as structural materials~\cite{Kruzic2016}, recent 
efforts have been focused on unveiling defects likely to determine mechanical 
behavior; in close analogy to well-known plastic flow mechanisms existing 
in crystalline systems. Particularly in this latter context, recent basic 
research has been conducted into atomic-scale characterization of structural 
heterogeneity~\cite{QiaoWang2019} and dynamic 'defects' from the perspective 
of \textit{flow units}~\cite{WangWang2018}. This is an aspect that reflects 
space-time heterogeneity in MGs and a technique that has been increasingly 
used to study the associated relaxation dynamics is dynamical mechanical 
analysis (DMA, aka dynamical mechanical spectroscopy), due to its high 
sensitivity in detecting atomic rearrangements~\cite{YuSamwer2013,YuWang2014,
QiaoPelletier2014,QiaoWang2017,YuRichert2017}. 

Still regarding structural characterization, large-scale molecular dynamics 
(MD) simulations have proven effective in providing trustworthy structural 
models of MGs; able to reproduce elementary properties like density, glass 
transition temperature, and representative statistics regarding atomic 
configurations~\cite{ChengMa2009}. Nevertheless, the quest for efficient 
strategies for handling realistic structural models (that often contain 
thousands of atoms) in such simulations is a problematic and topical 
issue~\cite{BartokDe2017}; even for the prediction of essential ``static'' 
features like short- and medium-range orders (SRO and MRO). In fact, this is 
a constraint that imposes a critical limitation on the use of \textit{ab 
initio} quantum mechanical MD simulations for that end; and the solution is 
nothing new: a multiscale approach, in which the role of electrons on 
interatomic interactions is abstracted and described in terms of the 
so-called interaction models (or interatomic potentials) employed in the 
well-known classical MD (CMD) simulations.

The central drawback of this approach, however, is that so many additional 
properties relying strongly on an accurate description of the material's 
electronic structure become simply not accessible in CMD simulations. In 
other words, having a thorough atomistic (structural topology) insight into 
a certain material may not be enough to reveal all the technologically 
relevant phenomena. A good example from recent literature is the key role 
played by local homopolar bonds on the structural stabilization of specific 
sites, and how it impacts the resistance-drift of amorphous phase-change 
materials~\cite{DeringerZhang2014}.

This article reports an unprecedented ``big picture'' view of chemical 
bond strengths in MGs of the Zr--Cu--Al (ZCA) alloy system. The link 
between electronic structure and first-principles chemical bonding 
information is given by density functional theory 
(DFT)~\cite{HohenbergKohn1964,KohnSham1965} and crystal orbital Hamilton 
population (COHP) analysis~\cite{DronskowskiBlochl1993}; whereas 
representative statistics is attained by applying a machine learning (ML) 
approach to realistic structural models generated by CMD simulations. 
Under the specific scope of mechanical loading, the resulting overview 
of chemical bond strengths revealed a chemical/structural heterogeneity 
that is quite in line with the tendency to \textit{bond exchange} verified 
for atomic local environments in the chosen alloy model system.

\section{Theory and Computational Details}

\subsection{The Machine Learning-Based Approach}
\label{ML}

The ZCA system was selected as a prototypical alloy for application of 
the proposed ML-based protocol because of its practicality. First of all, 
its corresponding MGs are conventional model systems extensively studied 
due to their high GFA~\cite{PaulyLiu2010}; and there are plenty of 
experimental and theoretical works available in the literature covering 
different topics on them. Moreover, the computational modeling process is 
made easy for this alloy given the availability of an interatomic 
potential~\cite{ChengMa2009} that has been widely used for years; providing 
valuable theoretical support to experimental studies until 
recently~\cite{WangInoue2018,PekinDing2019} (see Section S1 in 
the Supporting Information). For the goals of the present study, it certainly 
provides the required plausible description of ``static'' bulk SRO and MRO 
in these MGs with affordable 10000-atoms cells~\cite{ChengMa2009} -- naturally, 
assuming a homogeneous distribution of constituent elements in a glassy 
structure free of nanocrystals~\cite{JiaoLiu2017,KumarOhkubo2007}. 

The precise motivation for a ML-based approach here is to enable the 
prediction of bond strengths between atom pairs with the DFT accuracy in 
10000-atoms cells of the ZCA MGs derived from CMD simulations. Such cells 
are expected to supply representative statistics regarding SRO 
(chemical/structural local environments) in these systems. However, it 
is manifest that the corresponding electronic structure quantum mechanical 
simulations are unfeasible and unnecessary. This is where the proposed ML 
model comes onto the scene, by learning bond strengths from a database of 
interactions (DBIs).

Here, it is important to point a proper definition of chemical bond 
strength within the scope of this work. In fact, it can be seen as one 
more quantum mechanical concept associated to bond order -- i.e., the 
stability of a chemical bond indicated by the electron density preferably 
distributed within a region between the related pair of atoms rather 
than closer to the individual corresponding sites. Its origins lie in the 
linear combination of atomic orbitals (LCAO) molecular overlap population 
Mulliken formalism~\cite{Mulliken1955}, followed by its extension to solids 
(periodic extended systems) dubbed crystal orbital overlap population (COOP) 
analysis~\cite{HughbanksHoffmann1983}; which was built on top of a particular 
crystal orbital (CO) scheme combined with an extended H\"{u}ckel method. In 
a simple CO-tight-binding language, for two sites \textit{A} and \textit{B} 
with corresponding COs $\phi_{\nu}$ and $\phi_{\mu}$, the elements of the 
overlap population matrix in a closed shell system are 
\begin{equation*}
  P_{AB} = 2\sum_{\nu}^{A}\sum_{\mu}^{B}c_{\nu}^{*}c_{\mu}S_{\nu\mu};
  \label{eq:overlap_pop}
\end{equation*}
with $c_{\nu}^{*}c_{\mu}$ and $S_{\nu\mu}$ the elements of the density 
matrix and the overlap matrix, respectively. So, as in any CO approach, 
the COOP method is based on the density of states (DOS) curve (i.e, on 
the number of electronic states over the energy scale~\cite{Benco1995}) 
and, as an energy-resolved bonding descriptor, it provides information 
about the chemical bonding from a weighted DOS curve obtained by the 
product of $S_{\nu\mu}$ and DOS matrix elements in the applicable energy 
ranges.

The resulting COOP curve constitutes an electron number partitioning 
scheme and therefore carries features that allow the use of its integral 
up to the Fermi level (ICOOP) in close analogy to bond order as an index 
of bond strength; since the computed quantities $\Re\{c_{\nu}^{*}c_{\mu}S_{\nu\mu}\}$ 
point to bonding (positive), nonbonding (zero), and antibonding (negative) 
contributions~\cite{DronskowskiBlochl1993}. The COHP method (proposed 
in that same Ref.~\citenum{DronskowskiBlochl1993}) used in the present 
work is nothing but an alternative approach, in which the DOS curve is 
weighted by the Hamiltonian matrix elements ($H_{\nu\mu} = 
\bra{\phi_{\nu}}\hat{H}\ket{\phi_{\mu}}$) instead of the corresponding 
$S_{\nu\mu}$ employed in COOP. So, whereas the latter is said an 
electron-partitioning scheme, the former is an energy-partitioning 
scheme that minimizes the drawback caused by the basis-set dependence of 
the overlap integral and is very well suitable for first-principles DFT 
electronic structure simulations; but both are energy-resolved bonding 
descriptors.

Moving to the strategy for the generation of the database of structures 
to train the ML model (the aforementioned DBIs), it was grounded on the 
assumption that the referred statistical representativeness existing in 
a 10000-atoms cell of a given nominal composition (NC) of the ZCA alloy 
can be attained with a corresponding ensemble of smaller 100-atoms cells. 
As described in details in Section~\ref{ComputationalDetails}, all these 
cells were obtained by CMD simulated cooling from the melt; and the 
stated statistical equivalence was indeed achieved (see Supporting 
Information). Moreover, in order to draw a parallel with experiments 
reported in the literature~\cite{KumarOhkubo2007}, the set of four 
NCs (Zr$_{0.5}$Cu$_{0.5}$)$_{100-x}$Al$_x$ (with $x$ = 2, 6, 8, and 10) 
has been selected.

\begin{table}
 \caption{Overview in numbers of the per-type DBIs created for the MGs 
          using the 100-atoms cells generated following the quenching 
          protocol described in the text and whose amounts are indicated 
          in the last row. All cells are available in the 
          Supporting Information as \textit{extended XYZ} files.} 
  \label{tbl:DBIs}
  \begin{tabular}{crrr}
    \hline
    \multicolumn{1}{c}{} & 
    \multicolumn{1}{c}{Zr$_{49}$Cu$_{49}$Al$_{2}$} &
    \multicolumn{1}{c}{Zr$_{47}$Cu$_{47}$Al$_{6}$} &
    \multicolumn{1}{c}{Zr$_{45}$Cu$_{45}$Al$_{10}$}\\
    \hline
Al--Al          &     27 &    64 &   552\\
Cu--Al          &   5080 &  1775 & 17073\\
Cu--Cu          &  57230 &  6311 & 32859\\
Zr--Al          &   7011 &  2497 & 22938\\
Zr--Cu          & 169487 & 19153 & 99015\\
Zr--Zr          &  76919 &  8558 & 43547\\
    \hline
100-atoms cells &    488 &    58 &   326\\
    \hline
  \end{tabular}
\end{table}

Hence, a set of per-type DBIs has been created for all possible 
interaction types (ITs): Al--Al, Cu--Al, Cu--Cu, Zr--Al, Zr--Cu, and 
Zr--Zr. Table~\ref{tbl:DBIs} brings out an overview in numbers of 
these databases and details on their construction are provided in 
Section~\ref{ComputationalDetails}; including strategies to diversify 
the collection of chemical environments and to ensure transferability 
among different NCs. Here, it has to be pointed that the small number 
of Al--Al interactions shown in Table~\ref{tbl:DBIs} is intrinsic to 
the material. It follows upon alloying of Al and has already been 
described in the literature as a \textit{solute--solute avoidance} 
effect~\cite{ShengLuo2006,YuanYang2019}. As discussed in the Supporting 
Information, concerning the abovementioned transferability, this effect 
was not a complicating factor when further applying the ML model; and a 
single Al--Al DBI was set up by with all interactions of this type listed 
in Table~\ref{tbl:DBIs}.

Recalling that each individual interaction in the set of DBIs has an 
associated bond strength value derived from COHP analysis; precisely, this 
value is assumed to be the additive inverse of the integral of the COHP 
curve up to the Fermi level (-ICOHP). The minus sign is conventionally 
included to make it compatible with the ICOOP counterpart; since the system 
undergoes a lowering of its energy when there are bonding 
contributions~\cite{DronskowskiBlochl1993}. Hence, in the COHP curve, the 
product of $H_{\nu\mu}$ and the corresponding DOS matrix elements point to 
bonding (negative), nonbonding (zero), and antibonding (positive) values. 

Next, it is appropriate to introduce a key ``ingredient'' of the proposed 
ML model; whose versatility also allowed the assessment of the convergence 
of SRO statistics in the 100-atoms cells in advance. It refers to the 
mathematical descriptor of each chemical environment around an individual 
atoms; necessary to measure the ``distance'' (or dissimilarity) between atomic 
environments. The descriptor adopted in the present work is the smooth overlap 
of atomic positions (SOAP)~\cite{BartokKondor2013}. The SOAP fingerprints 
describe the chemical/structural local environments ensuring, in a natural 
way, invariance to the basic symmetries operations: rotation, reflection, 
translation, and permutation of atoms of the same species. In short, it 
describes the atomic neighborhood by expanding it in a basis composed of 
spherical harmonics and a set of orthogonal radial basis functions. The 
derived rotationally invariant power spectrum yield elements that are 
collected into a unit-length vector \textbf{q}~\cite{DeBartok2016}. So, in 
practice, the normalized similarity between such SOAP vectors computed for 
two atoms, $i$ and $j$, is given by their dot product, $\textbf{q}_i.\textbf{q}_j$.

Turning to the specification of the ML model, it is important to remark that 
its implementation is not intended to predict atomic (or per-atom) scalar 
quantities. The -ICOHP is a scalar property that is associated to atom 
pairs and, since the proposed ML model is founded upon a Gaussian process 
regression (GPR) framework, the function that measures the similarity 
between two chemical bonds (the \textit{kernel} function $k(B_m,B_n)$) has 
to take into account two basic features: bond distances and the individual 
SOAP vectors of the atoms involved. Thus, the covariance (or \textit{kernel}) 
function was defined as a squared exponential weighted by the normalized 
similarities given by the referred SOAP vectors, as following
\begin{widetext}
\begin{equation}
  k(B_m,B_n) = exp\left(-\frac{(d^m_{ij}-d^n_{ij})^2}{2\theta^2}\right)\left(\frac{(\textbf{q}^m_i.\textbf{q}^n_i+\textbf{q}^m_j.\textbf{q}^n_j+\textbf{q}^m_i.\textbf{q}^n_j+\textbf{q}^m_j.\textbf{q}^n_i)}{4}\right).
  \label{eq:kernel}
\end{equation}
\end{widetext}
With $d^m_{ij}$ the distance between atoms $i$ and $j$ in the chemical bond 
$B_m$; and $\textbf{q}^m_i$ and $\textbf{q}^m_j$ their corresponding SOAP 
vectors. The adjustable scaling parameter $\theta$ defines the ML model's 
behavior and sets the characteristic length-scale of the GPR. 

Finally, using the \textit{kernel} function in equation~\ref{eq:kernel}, the 
ML model is able to predict the -ICOHP value associated to an arbitrary bond 
$B$ according to
\begin{equation}
  ICOHP^{ML}(B) = \sum_{m=1}^{N}\alpha_mk(B,B_m)^\zeta.
  \label{eq:ICOHP}
\end{equation}
With the hyperparameter $\zeta$ = 1 establishing a linear \textit{kernel}; and 
$N$ the size of the training DBI at issue containing the set of reference chemical 
bonds $\left\{B_m\right\}_{m=1}^N$ (or training set) with their corresponding 
\textit{ab initio} -ICOHP values. By inverting the $N \times N$ \textit{kernel} 
matrix $\textbf{K}$, whose elements $K_{mn} = k(B_m,B_n)$ are defined with the 
training set, the per-interaction weights $\alpha_m$ in equation~\ref{eq:ICOHP} 
are computed as following
\begin{equation}
  \alpha_m = \sum_{n=1}^{N}\{\textbf{K}^\zeta + [(\sigma^2\gamma)\textbf{1}]\}^{-1}_{mn}ICOHP^{DFT}(B_n).
  \label{eq:alpha}
\end{equation}
With $\sigma^2$ the standard deviation of the -ICOHP values in the training set 
$\left\{B_n\right\}_{n=1}^N$; $\gamma$ a regularization adjustable parameter, 
$\textbf{1}$ the $N \times N$ unit matrix, and $ICOHP^{DFT}(B_n)$ the -ICOHP 
value of the reference chemical bond $B_n$ computed from first-principles DFT.

\begin{figure}[htb]
 \centering
 \includegraphics[scale=0.3]{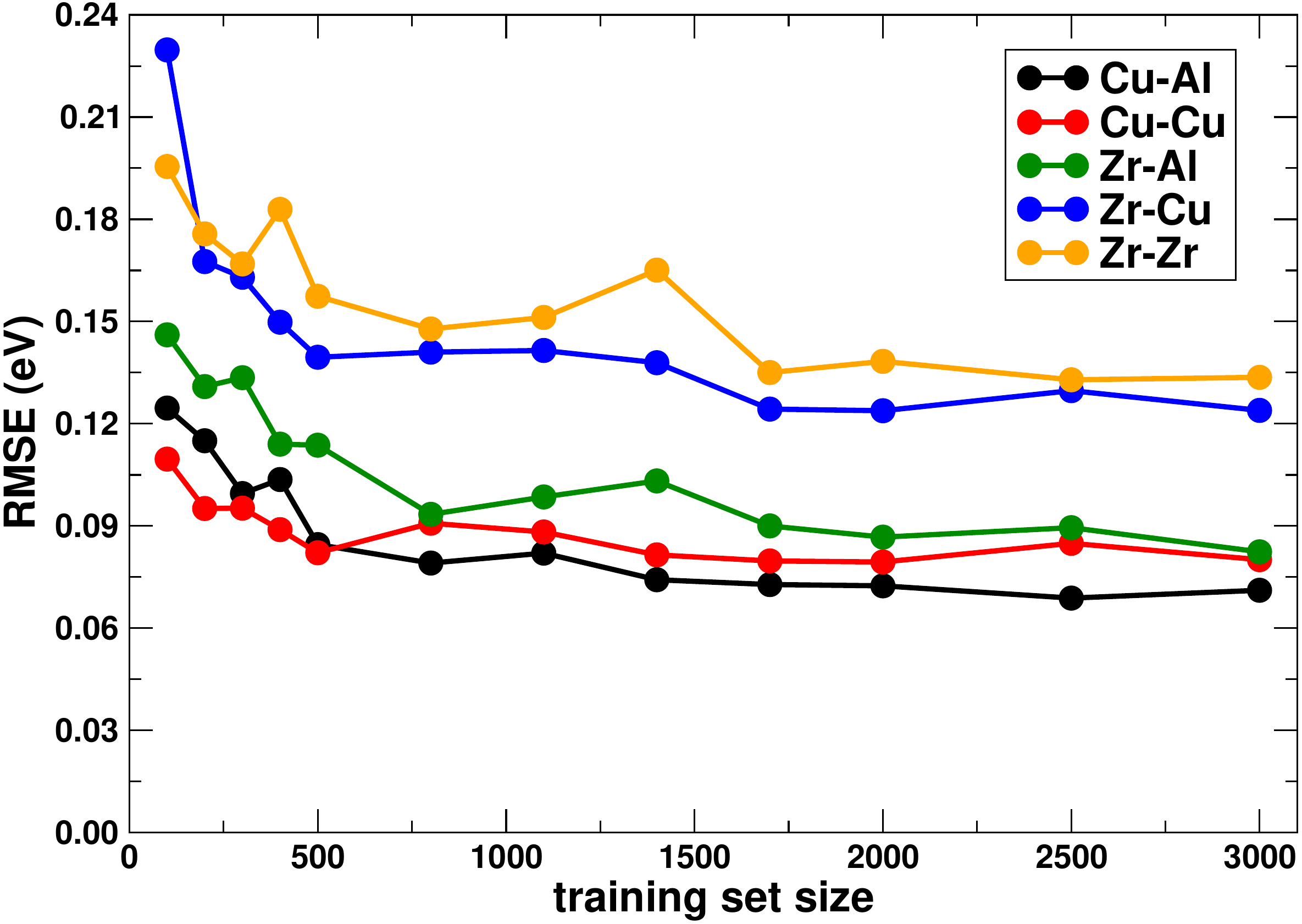}
 \caption{Root-mean-square errors (RMSE) calculated from different 
          Gaussian process regressions made for each interaction type 
          in the nominal composition Zr$_{45}$Cu$_{45}$Al$_{10}$ assuming 
          distinct training set sizes. The ML model parameters were set 
          as $\zeta$ = 1, $\theta$ = 0.5, and $\gamma$ = 0.010, and 
          a fixed testing set size of 1000 interactions was taken.}
 \label{img:RMSD}
\end{figure}

\subsection{Computational Details}
\label{ComputationalDetails}

All the CMD simulations were carried out using the velocity-Verlet integrator 
as implemented in the LAMMPS package~\cite{Plimpton1995} (release 16Feb2016). 
The embedded atom model (EAM) interatomic potential developed and properly 
tested by Cheng \textit{et al.}~\cite{ChengMa2009,Sheng2011} (see 
Section S1 in the Supporting Information) was used with a cutoff radius of 6.5 
\AA~to describe the interatomic forces in all MGs. A common quenching protocol 
was used for the generation of all 10000-atoms and 100-atoms cells introduced 
in Section~\ref{ML}. Firstly, an initial configuration was set by randomly 
positioning the atoms in a cubic supercell, whose initial volume was estimated 
for each NC by initially assuming a dense sphere packing weighted with the Zr, 
Cu, and Al atomic radii; with a length in excess of 10\% added to each cell 
vector. In order to avoid superposition of atoms, a conjugate gradient 
minimization on the random initial structure was executed with a stop criterion 
defined by a force threshold of 10$^{-8}$ eV/\AA. Next, the system was 
thermalized at 2000 K in the isothermal--isobaric (\textit{NPT}) ensemble for 2 
ns (in the time evolution of the CMD simulation ($\Delta$t)). With a time step 
of 2 fs (adopted in all runs), the Nos\'{e}-Hoover thermostat was used with a 
dump coefficient of 0.2 ps; whereas the barostat was set to zero pressure with 
a dump coefficient of 2 ps. Subsequently, the system was cooled to 300 K with 
a minimal and feasible rate of 8.5$\times$10$^9$ K/s; and finally, the glassy 
structure was allowed to relax at 300 K for $\Delta t$ = 2 ns.

As pointed out in Section~\ref{ML}, some of the 100-atoms cells listed in 
Table~\ref{tbl:DBIs} were submitted on demand to DFT first-principles calculations 
to set up the DBIs for the ML model. For the purpose of diversifying the collection 
of chemical environments, each one of those selected cubic supercells derived 
from CMD simulations was submitted to 3D geometrical transformations which have 
generated 14 new structures, namely: shearing along the x, y, and z axes (6 new 
structures); compression and tension along the x, y, and z axes (6 new structures); 
isotropic compression and tension (2 new structures).

All the referred DFT electronic structure simulations were performed using the 
{\sc Quantum} ESPRESSO~\cite{GiannozziBaroni2009,GiannozziAndreussi2017} (QE) 
open-source software suite version 6.2.0; with plane-wave (PW) basis sets and 
projector augmented waves (PAW)~\cite{Blochl1994} datasets from the PSLibrary 
project version 1.0.0~\cite{DalCorso2014}. The Perdew-Burke-Ernzerhof 
(PBE)~\cite{PerdewBurke1996} generalized gradient approximation was used to 
describe the exchange-correlation functional in all computations. Remarking 
that the abovementioned EAM interatomic potential~\cite{ChengMa2009,Sheng2011} 
used in the CMD simulations was also parametrized from results of DFT-PBE 
calculations. For all 100-atoms cells, the PW basis set was truncated with a 
kinetic energy cutoff of 70 Ry and the Monkhorst-Pack procedure~\cite{MonkhorstPack1976} 
was used to determine the k-points disposition in the first Brillouin zone 
from a 2$\times$2$\times$2 sampling (corresponding to a density of k-points 
of about 0.04 in all structures). A Fermi-Dirac probability distribution was 
used as a smearing function to set the occupations of energy levels, with a 
common broadening parameter $k_bT$ = 8 mRy. 

The ICOHP method has been introduced in Section~\ref{ML} and was earlier 
described in Ref.~\citenum{DronskowskiBlochl1993} as a scheme suitable for 
first-principles DFT calculations. However, further development was required 
to make it compatible with currently predominant and numerically efficient 
PAW-based computations with PW basis sets; what was achieved with the projected 
COHP (pCOHP) approach~\cite{DeringerTchougreeff2011,MaintzDeringer2013} which 
is implemented in the LOBSTER code~\cite{MaintzDeringer2016,MaintzEsser2016} 
version 3.0.0 employed in the present work. Apropos, this implementation has 
already proven to be effective when using PAW data generated by 
QE~\cite{NelsonKonze2017}; which are required for projections from delocalized 
wave-functions onto local auxiliary basis. For the referred projections, a 
default local basis set provided by Bunge \textit{et al.}~\cite{BungeBarrientos1993} 
was adopted with the following set of local orbitals for each atomic specie: 
Al ($3s~3p_x~3p_y~3p_z$); 
Cu ($3s~4s~3p_x~3p_y~3p_z~3d_{xy}~3d_{yz}~3d_{z^2}~3d_{xz}~3d_{x^2-y^2}$); 
and Zr ($4s~5s~4p_x~4p_y~4p_z~4d_{xy}~4d_{yz}~4d_{z^2}~4d_{xz}~4d_{x^2-y^2}$). 
Here, it is important to report that the high quality and reliability of these 
projections is reflected by the small values of the charge spilling (no higher 
than 1.6\%) in all ICOHP calculations with 100-atoms cells reported in this 
work -- in passing, absolute total spilling did not exceed 4.8\%.

The SOAP descriptors were generated using the QUIP package~\cite{Gabor2019}; 
with a cutoff of 3.75 \AA~for the definition of the range of each chemical 
environment around atoms, including all elements in its composition. This 
cutoff value was based on partial pair distribution functions that are 
well-known for the MGs of the ZCA alloy~\cite{ChengMa2009}. The spherical 
harmonics basis band limit and the number of radial basis functions were set 
to 6 and 8, respectively; with all the remaining parameters kept with their 
corresponding default values.

\section{Results and Discussion}

\subsection{Predictive Power of the ML Model}

The predictive power of the ML model given by equation~\ref{eq:ICOHP} was 
evaluated individually for each interaction type (IT) by studying its 
convergence with respect to the training set size, for a fixed testing set 
size. Due to its completeness (see Table~\ref{tbl:DBIs}), the set of DBIs 
created for the NC Zr$_{45}$Cu$_{45}$Al$_{10}$ was selected for initial tests; 
for which the ML model parameters were arbitrarily set as $\zeta$ = 1, 
$\theta$ = 0.5, and $\gamma$ = 0.010 (see also equations~\ref{eq:kernel} 
and~\ref{eq:alpha}) and the corresponding results are presented in 
Fig.~\ref{img:RMSD}. From which it is possible to see that the corresponding 
root-mean-square errors (RMSE) are dependent on the IT but, in general, 
they are fairly converged with small training set sizes (from about 500 
interactions). Furthermore, given the range of -ICOHP values computed for 
the MGs under study listed in Table S1, the individual RMSE values calculated 
for each IT show that the ML model produces small prediction errors. Remarking 
that each point in the plots of Fig.~\ref{img:RMSD} corresponds to a GPR 
in which the training and testing sets were randomly created from the DBIs 
listed in Table~\ref{tbl:DBIs}.

\begin{figure}[htb]
 \centering
 \includegraphics[scale=0.33]{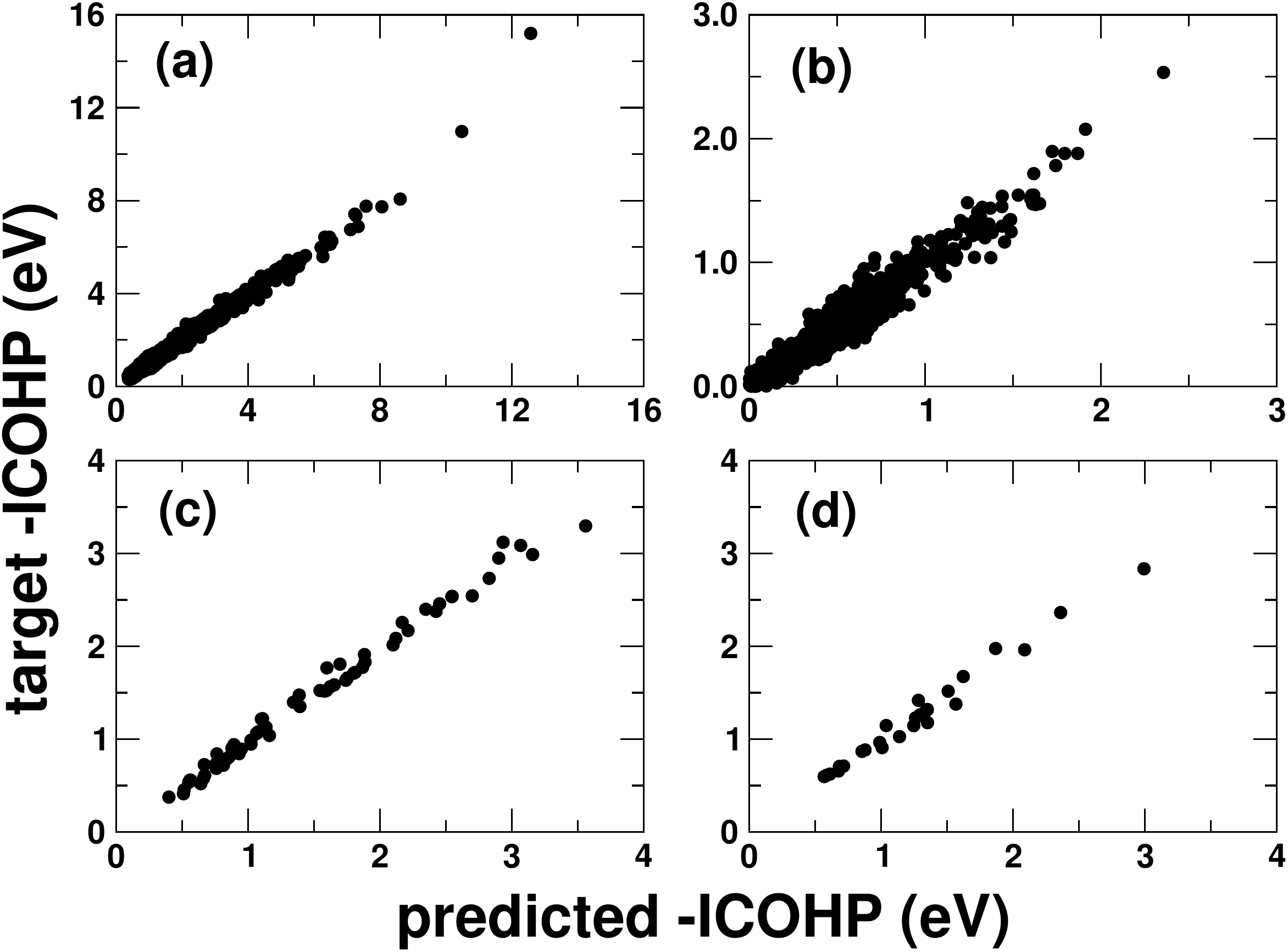}
 \caption{Scatter plots of results from some of the Gaussian process 
          regressions listed in Table S1: 
          (a) Zr--Zr in Zr$_{45}$Cu$_{45}$Al$_{10}$; 
          (b) Cu--Cu in Zr$_{45}$Cu$_{45}$Al$_{10}$; 
          (c) Al--Al in Zr$_{47}$Cu$_{47}$Al$_{6}$; and 
          (d) Al--Al in Zr$_{49}$Cu$_{49}$Al$_{2}$.}
 \label{img:scatter}
\end{figure}

Table S1 provides further statistics from a set of GPRs equivalent to those 
shown in Fig.~\ref{img:RMSD}; however adopting rather large training and 
testing sets. As already mentioned, the lack of Al--Al interactions is 
intrinsic to the ZCA alloy and made it impossible to carry out the same 
thorough tests for that specific IT. Nevertheless, it can be checked in 
Table S1 that the corresponding RMSE values also point to small prediction 
errors. With respect to the other two NCs Zr$_{47}$Cu$_{47}$Al$_{6}$ and 
Zr$_{49}$Cu$_{49}$Al$_{2}$, it can be verified in Tables S2 and S3 that 
similar RMSE values were found for their correlated DBIs. Hence, since 
RMSE values are scale-dependent, the overall predictive power of the 
proposed ML model can be said satisfactory within the context of this study. 
The scatter plots of some particular ITs listed in Table S1 are shown in 
Fig.~\ref{img:scatter}; and the complete list is available in the Supporting 
Information (Figs. S13 to S30).

\subsection{Application of the ML Model}

\subsubsection{Static Structures at Room Conditions}
\label{ApplicationStatic}

Once the validation is complete, the ML-based approach was used to predict 
bond strengths between atom pairs in the four 10000-atoms cells created for 
each NC of the ZCA alloy. Based on the convergence tests and transferability 
of individual DBIs explained in the Supporting Information, a minimal and 
feasible training set ($\left\{B_n\right\}_{n=1}^N$) with 600 interactions 
for each IT has been set by merging equally the corresponding DBIs listed in 
Table~\ref{tbl:DBIs}. The referred 10000-atoms cells are able to provide the 
required statistical representativeness regarding chemical environments in 
these MGs (see Section S1 in the Supporting Information) and they were 
generated with CMD simulations following the protocol described in 
Section~\ref{ComputationalDetails}.

\begin{figure*}[htb]
 \centering
 \includegraphics[scale=0.6]{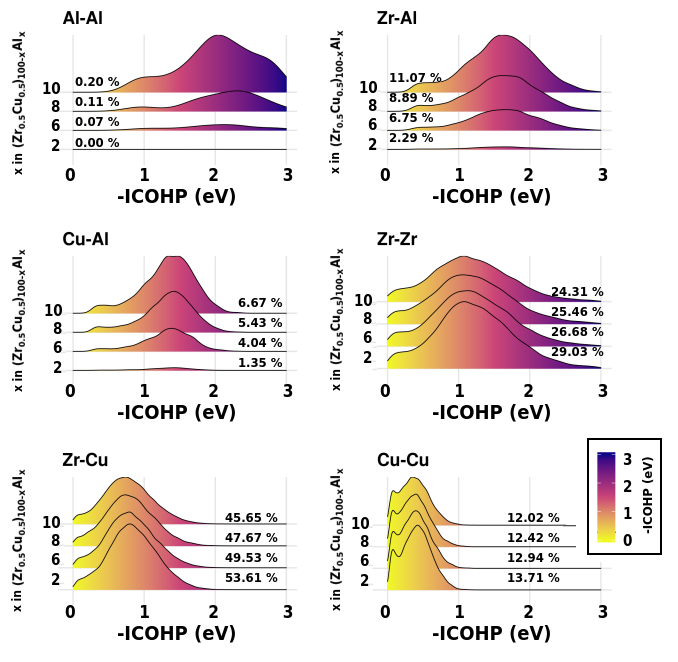}
  \caption{Distributions of -ICOHP values predicted by the ML model 
          for all interatomic interactions existing in 10000-atoms 
          cells of the series of nominal compositions (NC) 
          (Zr$_{0.5}$Cu$_{0.5}$)$_{100-x}$Al$_x$ (with $x$ = 2, 6, 8, 
          and 10). The fractions of each interaction type (IT) in each 
          NC are indicated as percentages.}
 \label{img:ICOHP_ZCA-x}
\end{figure*}

\begin{figure*}[htb]
 \centering
 \includegraphics[scale=0.55]{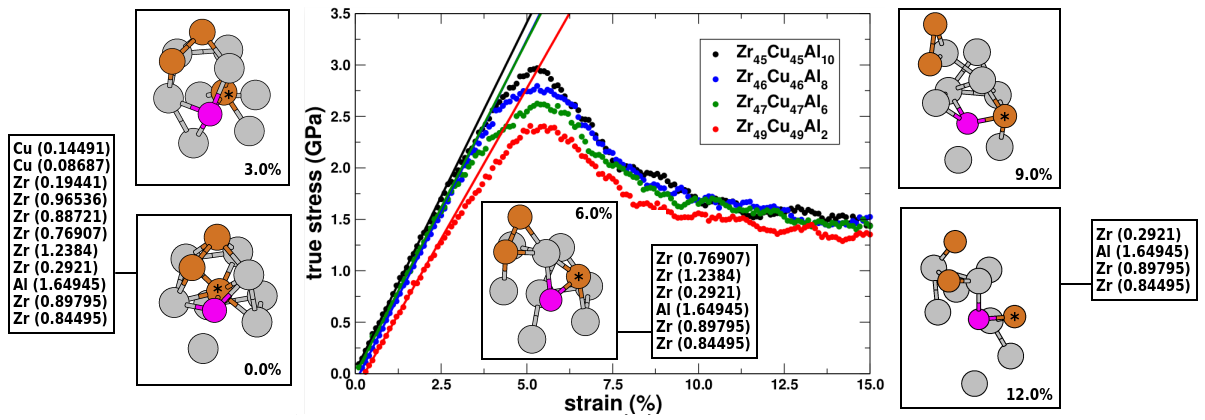}
  \caption{Stress-strain curves simulated at 300 K with a compression rate of 
          1$\times$10$^7$ s$^{-1}$ for the MGs studied in this work. The solid 
          lines denote the elastic behavior (see also Fig. S32). The time 
          evolution of a specific Cu-centered (marked with an asterisk) non-persistent 
          local environment (NPLE, see text) extracted out of the structural model 
          of the MG Zr$_{45}$Cu$_{45}$Al$_{10}$ is depicted; whose snapshots are 
          labeled with the corresponding strain value. The -ICOHP values (in eV) 
          of the first-neighbours computed for the 0\% strained structure are 
          listed. These same values are repeated (not recalculated) for 
          the persistently bonded atoms in the 6\% and 12\% strained NPLEs. Zr, 
          Cu, and Al atoms are shown in grey, brown, and pink colors, respectively.}
 \label{img:ss_w}
\end{figure*}

Above all, this work is aimed at delivering an approach able to promote 
insights into the interplay of chemical/structural and dynamical 
inhomogeneities existing in MGs in different contexts -- to be explored in 
future studies. The most promising applications comprise research focused 
on topical conceptions in which atom categorization based on the strength 
of chemical bonds plays a key role; like the aforementioned atomic models 
in terms of \textit{flow units}~\cite{WangWang2018} or \textit{tightly 
bonded clusters}~\cite{FanLiaw2009}, \textit{weakest 
configurations}~\cite{LiJiang2017}, \textit{bond exchange}~\cite{JiaoWang2015} 
processes, among others. Therefore, instead of going straight to the 
issue of how -ICOHP values are 3D-distributed within the 10000-atoms 
cells, this work will be limited to the assessment of the propensity for 
\textit{bond exchange} of individual atomic local environments under the 
scope of mechanical loading; and then, evaluate its connection to bond 
strength and atomic mobility using ordinary descriptive statistics. 
However, before doing so, it is worth looking at a simpler per-bond 
analysis as depicted in Fig.~\ref{img:ICOHP_ZCA-x}. 

Most importantly, it should be emphasized that the set of results in 
Fig.~\ref{img:ICOHP_ZCA-x} represents an unprecedented ``big picture'' 
view of bond strengths between atoms in these MGs derived from quantum 
mechanics. With respect to chemical bonding information, that comprehensive 
overview is arguably more complete and revealing than previous reports 
with similar aims; including quantum chemistry calculations with localized 
basis sets for rather small isolated atomic 
clusters~\cite{LekkaEvangelakis2009,LekkaBokas2012}; or even \textit{ab 
initio} MD simulations carried out with periodic boundary conditions, 
but for cells containing a few hundred interatomic interactions~\cite{ChengMa2009}.

In fact, the profiles of the distributions in Fig.~\ref{img:ICOHP_ZCA-x} 
do not exhibit significant changes upon alloying of Al and, as expected, 
only their respective fractions vary distinctively in each IT (indicated 
as percentages). Yet, they suggest that atoms can be grouped according to 
their chemical bonding situations; pointing to the possibility of drawing 
strategies to segregate atoms in a 3D fashion based on a given reference 
-ICOHP value. This is reasonable indeed, and one can see that Cu--Cu 
interactions show up as the weakest ones, together with a non-negligible 
fraction of Zr--Cu and Zr--Zr bonds; all having -ICOHP values below a value 
around 0.7 eV. Conversely, substantial amounts of relatively strong Zr--Cu 
and Zr--Zr bonds are also predicted by the ML model. 

One further remark concerning the Al alloying effect is that, despite 
the negligible number of Al--Al bonds -- related to the already introduced 
\textit{solute--solute avoidance} effect~\cite{ShengLuo2006,YuanYang2019} -- 
a fair amount of strong Zr--Al and Cu--Al interactions with -ICOHP 
$\gtrapprox$ 1.0 eV rise. And based on the percentages of ITs pointed out 
in Fig.~\ref{img:ICOHP_ZCA-x}, it can be inferred that these two latter are 
replacing the weaker Zr--Zr and Zr--Cu bonds; given the small variation of 
the weakest Cu--Cu bonds. This is likely to be the main chemical effect of 
Al alloying, which is reflected in the quite distinctive mechanical behavior 
upon mechanical loading reported for the two extreme NCs reported in the 
literature~\cite{KumarOhkubo2007}: whereas the MG Zr$_{45}$Cu$_{45}$Al$_{10}$ 
was found to be very brittle, the lowest Al-content counterpart 
Zr$_{49}$Cu$_{49}$Al$_{2}$ presented a large plasticity that has been 
assigned to localized nanocrystallization in the structure upon mechanical 
loading. This is the idea that will be further explored in next section; 
through a straightforward assessment of the propensity for \textit{bond 
exchange} of individual atomic local environments.

\subsubsection{Structures Subjected to Uniaxial Compression}

So far, the distributions in Fig.~\ref{img:ICOHP_ZCA-x} do not express any 
information concerning the collective role of the set of bonds existing in each 
individual local environment (LE, i.e., the neighborhood surrounding a central 
atom). Moreover, the respective bond strengths were extracted out of snapshots 
of static structures equilibrated at room environment conditions; what is 
not enough to assess the propensity for \textit{bond exchange}, which 
is a dynamical aspect of the MGs under study. So, in order to account these 
factors under the specific scope of uniaxial compression, a set of additional 
CMD simulations have been planned.

Stressing that the ultimate goal of these extra simulations is to show that 
there is a consistency between the chemical bonding heterogeneity disclosed 
by the ML model (as shown in Fig.~\ref{img:ICOHP_ZCA-x}), and the time 
evolution of the structural topology of individual LEs upon mechanical loading 
-- as described by the Newton's equations and underlying interaction model 
(the EAM potential). And it is important to bare in mind that these are two 
distinct aspects of the chemistry and dynamics, respectively, that are being 
revealed in this work through two completely independent computational 
approaches; whose complementarity has a great potential to bring insights 
into the synergic role played by chemical/structural and dynamical 
heterogeneities in MGs.

It was introduced in Section~\ref{ML} that the set of four NCs 
(Zr$_{0.5}$Cu$_{0.5}$)$_{100-x}$Al$_x$ (with $x$ = 2, 6, 8, and 10) has been 
selected aiming at drawing a parallel with experimental results reported in 
the literature~\cite{KumarOhkubo2007}; and the choice of that experimental 
study was motivated by two key aspects. First of all, the samples were 
subjected to uniaxial compression; a mechanical load test with a constrained 
geometry that simplifies its computational modeling, providing reliable 
results for the evaluation of the time evolution of LEs, as sought in this 
work. Additionally, the authors resorted to transmission electron microscopy 
and 3D atom probe tomography to characterize the microstructure of the samples; 
what provided the proper experimental backing to describe them as homogeneous 
amorphous alloys (i.e., free of nanocrystals) -- at least before compression. 
This is quite relevant regarding computational modeling, since that is the 
precise microstructure represented by the 10000-atoms cells used as structural 
models in the present work.

So, the corresponding CMD simulations were carried out for all NCs (with a 
rather tight time step of 1 fs) and they are quite consistent -- further 
technical details are provided in the Supporting Information in a 
comprehensive way and complementary to those introduced in 
Section~\ref{ComputationalDetails}. However here, it is worth commenting 
that MD simulations of uniaxial compression require proper approximation 
to describe temperature and pressure dissipation. In this work, the equations 
of motion proposed by Melchionna \textit{et al.}~\cite{MelchionnaCiccotti1993} 
have been employed with periodic boundary conditions in the \textit{NPT} 
ensemble; in which temperature is controlled with the Nos\'{e}-Hoover 
thermostat, whereas pressure is regulated by decoupling the boundary in 
the loading direction from the \textit{NPT} equations governing the other 
two orthogonal directions. This approach has already proven effective 
elsewhere~\cite{TschoppMcDowell2008}, and was fairly satisfactory for the 
purposes of the present study.

Furthermore, for the sake of transparency, there is another technical 
aspect of these CMD simulations that deserves a critical remark here in 
the main text; and it is not commonly discussed in the literature. The 
accurate atomistic simulation of the stress-strain (SS) curves measured 
in the experiments reported by Kumar \textit{et al.}~\cite{KumarOhkubo2007} 
requires structural models able to describe concomitantly two key facets: 
the role of surface effects on the brittle behavior of the MG 
Zr$_{45}$Cu$_{45}$Al$_{10}$; and the localized nanocrystallization 
(particles with some nanometers) taking place within shear bands, as well 
as its relationship with the extended plasticity reported for the MG 
Zr$_{49}$Cu$_{49}$Al$_{2}$.

The solution could be to increase the structural models up to some dozens 
(or even few hundreds) of millions of atoms; or, in a much more consistent 
approach, resort to a multiscale-based strategy. For instance, one could 
derive an upscaled peridynamic model to describe the brittle behavior of 
the former MG; or parameterize a constitutive model to describe the SS curve 
of the latter MG accounting for localized nanocrystallization. However, this 
is definitely out of the scope of the present work and unnecessary for its 
aims. That issue is well discussed in Section S2 in the Supporting 
Information. In short, the SS curves depicted in Fig.~\ref{img:ss_w} fairly 
reproduce the linear elastic behavior of the MGs; however, the portions that 
correspond to yield strength (strain $\approx$ 5\%) and further plastic 
deformation (strain $>$ 5\%) bear two common artifactual features 
that will certainly not affect further discussion within the context of the 
aims of the present study. Nevertheless, from here onward, most of the 
analyses will be restricted to strain $\lessapprox$ 6\%; although the extended 
segments of the SS curves will be carefully taken into account for examining 
particular aspects of the LEs; which are unrelated to the macromechanics of 
the corresponding MGs.

Returning to the topic of the time evolution of LEs in the MGs upon mechanical 
loading, in addition to the SS curves, Fig.~\ref{img:ss_w} also brings an 
example of a particular Cu non-persistent local environment (NPLE) extracted 
out of the structural model of the MG Zr$_{45}$Cu$_{45}$Al$_{10}$; whose central 
atom is marked with an asterisk. First of all, in the terminology of this work, 
a NPLE is a LE whose neighborhood has changed permanently over the mechanical 
loading -- i.e., through some kind of migration or rearrangement mechanisms, it has 
underwent \textit{bond exchanges} and some of its neighbor atoms have been 
replaced. Such changes were monitored in all CMD simulations for the two 
extreme NCs Zr$_{45}$Cu$_{45}$Al$_{10}$ and Zr$_{49}$Cu$_{49}$Al$_{2}$; and the 
referred example Cu-centered NPLE has lost five of its eleven neighbours at 6\% strain. 
Pointing that such NPLE broken bonds will be referred to as NPLE-BBs in this work; 
whereas, in contrast, those remaining persistent bonds will be dubbed NPLE-PBs. 

Additionally, it is self-evident that, for illustrative purposes, new neighbours 
are not shown in Fig.~\ref{img:ss_w}; and the neighbor's -ICOHP values 
predicted by the ML model for the non-strained structure (those from 
Fig.~\ref{img:ICOHP_ZCA-x}) are shown, primarily, to serve as a label for 
identifying the corresponding atoms in further strain values -- i.e., they have 
not been recalculated in the new strained structures. Moreover, in order to 
circumvent thermal fluctuation effects when monitoring neighbours, an additional 
of 0.50 \AA~has been added to the cutoff value of 3.75 \AA~introduced in 
Section~\ref{ComputationalDetails}; and this is the only arbitrary parameter used 
in this procedure.

The Cu-centered NPLE in Fig.~\ref{img:ss_w} provides just an extreme example of a set 
of \textit{bond exchange} processes in which the central atom has lost 64\% of 
its original neighbours at (``spurious'') 12\% strain. From that figure, it can 
be seen that the LE of this atom has indeed underwent severe changes over 
compression; and, as expected, this is a Cu atom whose bonds are overall week 
and hence representative of the corresponding distributions depicted in 
Fig.~\ref{img:ICOHP_ZCA-x}. The complete list of bond strengths together with 
the respective bond distances of this particular NPLE is available in Table S4; 
where it is noticeable that, incidentally, all the broken bonds have an 
associated -ICOHP value that is less than 1.0 eV. Naturally there are exceptions, 
since the dynamics of such \textit{bond exchange} processes is influenced by 
multiple factors like the number of nearest neighbours, relative distribution 
of the respective bond strengths around the coordination sphere, vibration 
fluctuations and related entropic effects, among others. 

So, in order to provide a complementarity support to that line of reasoning, 
a second contrast example has been extracted out of the same 10000-atoms cell 
of the MG Zr$_{45}$Cu$_{45}$Al$_{10}$. It refers to a Zr-centered NPLE that 
has exchanged only 26\% of its bonds at 12\% strain. The associated bonding 
data is also listed in Table S4 and the snapshots of its time evolution over 
compression are depicted in Fig. S37. That second example of NPLE reinforces 
the idea that all those factors influencing the dynamics of \textit{bond 
exchange} processes mentioned in last paragraph must be taken into account 
if one aims at specifying a type of local descriptor, able to allow sound 
segregation of atoms in MGs. In fact, atom categorization is a very promising 
application of the ML-based approach proposed in this work; capable of opening 
up new prospects and of playing a valuable role in further studies focused 
on current conceptions regarding chemical/structural and dynamical 
inhomogeneities existing in these materials other than \textit{bond 
exchange}~\cite{JiaoWang2015} processes. Nevertheless, the development of 
such referred descriptors definitely goes beyond the goal of the present work.

Moving to descriptive statistics of the whole 10000-atoms cells of the two 
extreme NCs Zr$_{45}$Cu$_{45}$Al$_{10}$ and Zr$_{49}$Cu$_{49}$Al$_{2}$, 
before all, the reader is referred to Figs. S38 and S39 to see that extreme 
cases like the Cu-centered NPLE shown in Fig.~\ref{img:ss_w} are rare in the 
CMD simulations. At 6\% strain, most of the NPLEs have no more than 10\% of 
broken bonds -- i.e., NPLEs-BBs $\lessapprox$ 10\% for most of the cases and 
that is a quantity that indicates the extent of \textit{bond exchange} in 
individual NPLEs. Also from Figs. S38 and S39, it is possible to verify that 
even at (``spurious'') 12\% strain, NPLEs-BBs $\gtrapprox$ 50\% are unusual.

\begin{figure}[htb]
 \centering
 \includegraphics[scale=0.3]{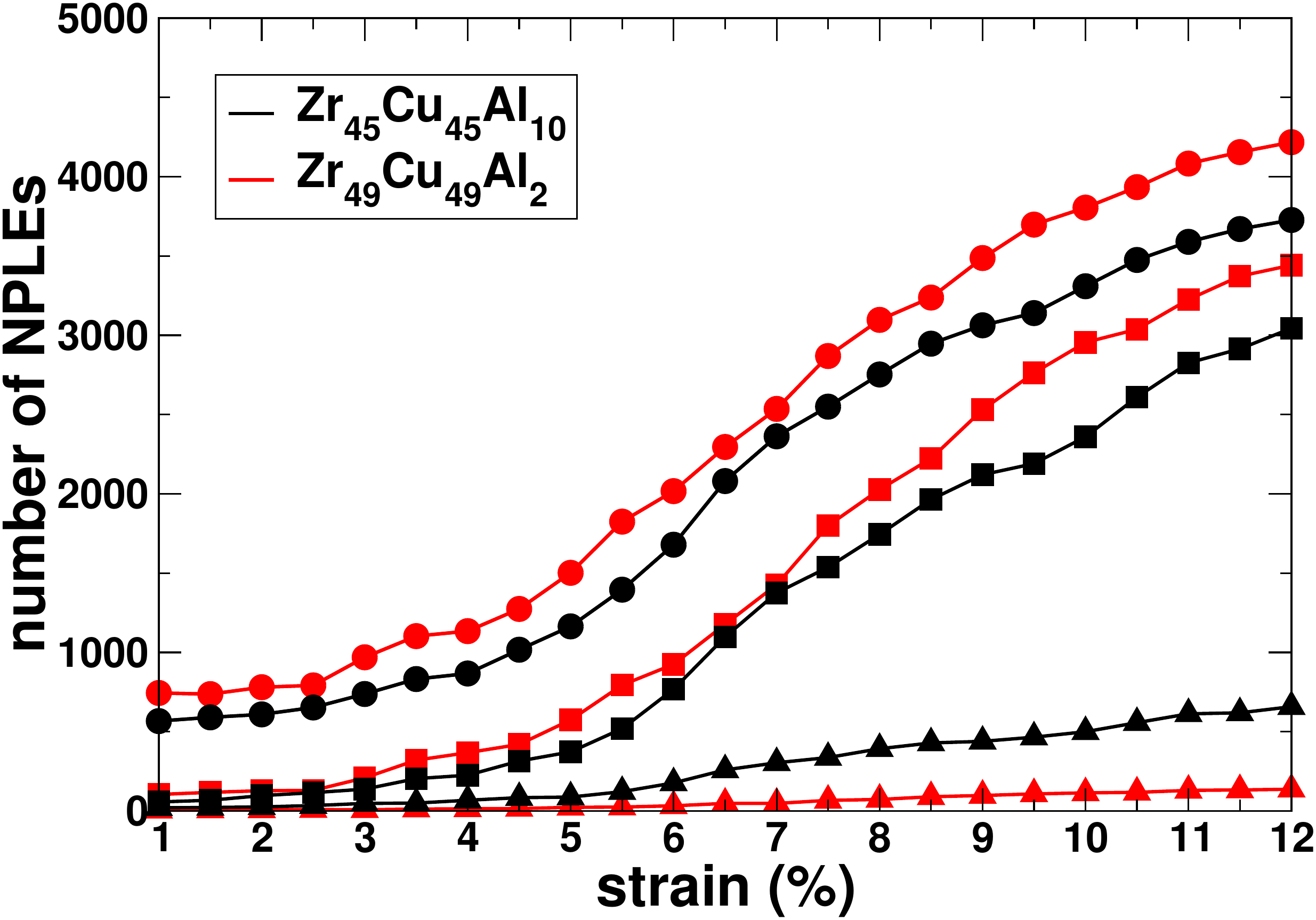}
 \caption{Time evolution over compression of the number of non-persistent local 
          environments (NPLEs, see text) centered at Zr ($\mycircle{black}$), 
          Cu ($\mysquare{black}$), and Al ($\mytriangle{black}$) atoms; counted 
          for the two extreme NCs Zr$_{45}$Cu$_{45}$Al$_{10}$ and 
          Zr$_{49}$Cu$_{49}$Al$_{2}$.}
 \label{img:SS-NPLEs}
\end{figure}

\begin{figure*}[htb]
 \centering
 \includegraphics[scale=0.6]{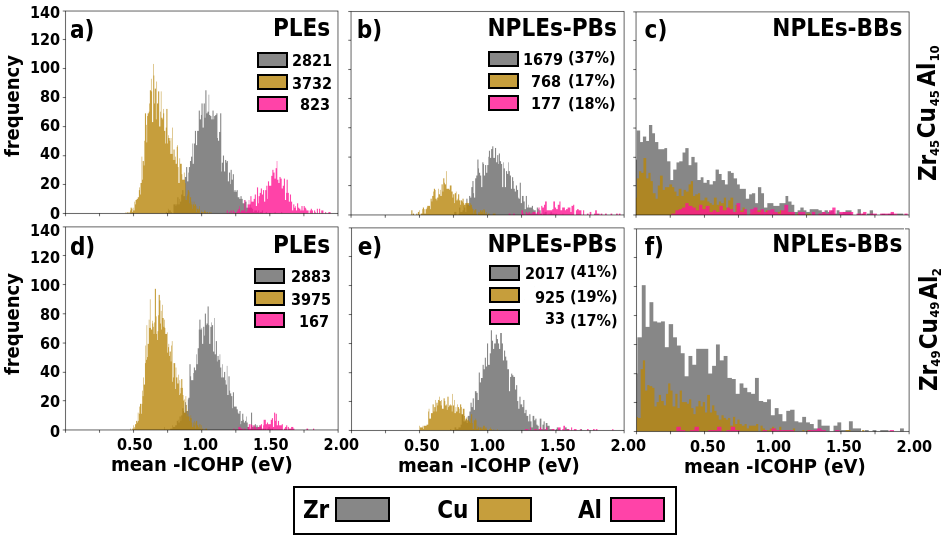}
  \caption{Distributions of mean -ICOHP values of persistent bonds (PBs) and 
           broken bonds (BBs) in persistent local environments (PLEs) and in 
           non-persistent local environments (NPLEs, see text) existing in the 
           10000-atoms cells of the two extreme NCs Zr$_{45}$Cu$_{45}$Al$_{10}$ 
           and Zr$_{49}$Cu$_{49}$Al$_{2}$. The number of bins in each histogram 
           was calculated from Sturges's formula and arbitrarily multiplied 
           by 18; whereas the corresponding colors are defined according to the species 
           of central-atoms. The total numbers of PLEs and NPLEs are shown; with 
           the percentages of NPLEs also provided between brackets.}
 \label{img:histmeans}
\end{figure*}

With respect to the amounts of NPLEs, Fig.~\ref{img:SS-NPLEs} brings the 
corresponding per-specie time evolution over compression. It goes without 
saying that those bare numbers reflect the stoichiometries of each NC, but 
it is clear that Zr atoms show higher propensity of bond breaking as already 
reported in the literature~\cite{JiaoWang2015}. Moreover, it is possible 
to see in Table S5 that at 6\% strain, the amounts of Zr- and Cu-centered 
NPLEs are relatively greater in the lowest Al-content NC 
Zr$_{49}$Cu$_{49}$Al$_{2}$ (in percentage terms regarding individual 
stoichiometries). For instance, whereas 41\% of the Zr atoms are NPLEs in 
that latter NC, this number drops to 37\% in the highest Al-content MG 
Zr$_{45}$Cu$_{45}$Al$_{10}$; and in the case of Cu atoms, that reduction 
is from 19\% to 17\%, respectively. 

From the counting results in Fig.~\ref{img:SS-NPLEs}, it can be said, in 
principle, that the propensities for \textit{bond exchange} of solvent Zr 
and Cu atoms -- in the CMD simulations carried out in the present work -- 
decrease upon alloying of Al; and this deduction has to be linked to the 
drastic change in the experimental mechanical behavior of the same MGs 
reported by Kumar \textit{et al.}~\cite{KumarOhkubo2007}. Indeed, one can 
see in Fig.~\ref{img:ss_w} that the onset of plastic deformation takes 
place earlier in the MG Zr$_{49}$Cu$_{49}$Al$_{2}$; which is the plastic 
NC in Ref.~\cite{KumarOhkubo2007}. Moreover, it is also noticeable in 
Fig.~\ref{img:ss_w} that the deviation of the linear elastic behavior 
occurs first and is more pronounced in that lowest Al-content NC than in 
the MG Zr$_{45}$Cu$_{45}$Al$_{10}$ (the one found to be brittle in 
Ref.~\cite{KumarOhkubo2007}). This is of course in line with the common 
perception that local bonds in that latter NC are less susceptible to 
be broken; since they become more strong, and maybe more covalent in 
addition to ionic contribution (i.e., more uni-directional and less 
isotropic than the metallic counterparts).

In fact, none of this is new. However, the actual missing information 
regarding bond strengths is the semiquantitative quantum mechanical analysis 
provided by the COHP method~\cite{DronskowskiBlochl1993}; what is being 
enabled by the ML model proposed in this work. Naturally, before the 
universe of thousands of predicted -ICOHP values depicted in 
Fig.~\ref{img:ICOHP_ZCA-x}, the analysis of individual cases like the Cu- 
and Zr-centered NPLEs discussed above will not offer the sought assessment of the 
propensity for \textit{bond exchange} under the scope of mechanical loading. 
So in order to accomplish this, in a word, Fig.~\ref{img:histmeans} brings 
a combination of the bonding information from Fig.~\ref{img:ICOHP_ZCA-x} 
with the outcomes of the CMD simulations depicted in Figs.~\ref{img:ss_w} 
and~\ref{img:SS-NPLEs}.

The histograms in Fig.~\ref{img:histmeans} represent the distributions of 
mean -ICOHP values computed for the zero strain structures (those from 
Fig.~\ref{img:ICOHP_ZCA-x}) for the two extreme NCs 
Zr$_{45}$Cu$_{45}$Al$_{10}$ and Zr$_{49}$Cu$_{49}$Al$_{2}$. However, in 
order to keep track on \textit{bond exchange} processes upon mechanical 
loading (as done for the Cu-centered NPLE in Fig.~\ref{img:ss_w}), all the 
histograms correspond to LEs existing in the 6\% strained structures. So, 
those per-bond -ICOHP values from Fig.~\ref{img:ICOHP_ZCA-x} have been cast 
into a per-atom (or per-LE) representation; dividing the neighbours of each 
central atom in three distinct groups, from which the referred mean bond 
strengths have been computed.

The former group comprises the neighbours of atoms whose LEs have been 
preserved throughout the entire uniaxial compression in the CMD simulations 
(up to 12\%); those are labeled PLEs (an abbreviation for persistent LEs) 
and they make up the histograms in Figs.~\ref{img:histmeans}(a) and (d). 
The other two groups comprehend the already introduced NPLEs; however, they 
are evaluated using two distinct sets of histograms for the same entries, 
namely: the persistent bonds (NPLEs-PBs) and the broken bonds (NPLEs-BBs) 
-- it refers to the histograms in Figs.~\ref{img:histmeans}(b), (c), (e), 
and (f). For example, the same Cu-centered NPLE depicted in Fig.~\ref{img:ss_w} 
at 6\% strain is an entry of the corresponding histograms in 
Figs.~\ref{img:histmeans}(b) and (c); with the corresponding arithmetic 
mean -ICOHP values averaged over its six PBs and its five BBs, respectively.

Back to the matter of the drastic change in the mechanical behavior of the 
MGs Zr$_{49}$Cu$_{49}$Al$_{2}$ (plastic) and Zr$_{45}$Cu$_{45}$Al$_{10}$ 
(brittle) reported by Kumar \textit{et al.}~\cite{KumarOhkubo2007}; this 
experimental outcome can be revisited in the light of the histograms from 
Fig.~\ref{img:histmeans}. First of all, they show that in all LEs (persistent 
or not) the means computed for persistent bond strengths (PLEs and NPLEs-PBs) 
are very dependent on the chemical specie of the central atom; regardless of 
the composition of its first coordination shell (i.e., the chemical species of 
neighbor atoms). Moreover, the profiles of the corresponding distributions 
also do not depend on the NC of the studied MGs. As can be seen, for all 
Zr-centered PLEs and NPLEs-PBs, the mean -ICOHP values are distributed 
around about 1.1 eV; whereas the corresponding distributions for Cu- and 
Al-centered LEs are centered around about 0.8 eV and 1.5 eV, respectively. 
Moreover, it is noticeable that the range of mean -ICOHP values covered by 
those histograms for Zr- and Cu-centered LEs (between 0.5 and 1.3 eV) cover 
the same range of most of the moderately stronger Zr--Zr and Zr-Cu bonds as 
shown in Fig.~\ref{img:ICOHP_ZCA-x}; what allows pointing 0.8 eV as a rough 
minimal value for what can be stated as a persistent bond in these materials. 

Still regarding persistent bonds, the differences concerning the Al-content 
of the two ZCA alloys lie mostly in the frequencies of observations; markedly 
those associated to Zr-centered NPLEs-PBs -- the grey histograms in 
Figs.~\ref{img:histmeans}(b) and (e). Pointing that these histograms are 
not normalized, as well as the counting results in Fig.~\ref{img:SS-NPLEs}. 
Despite this, the referred difference is still evident; what suggest that 
the drop from 41\% to 37\% of Zr-centered NPLEs caused by alloying of Al is 
a key factor related to the embrittlement of the MG Zr$_{45}$Cu$_{45}$Al$_{10}$, 
with respect to the lowest Al-content counterpart. However, especially 
concerning the plasticity of the MG Zr$_{49}$Cu$_{49}$Al$_{2}$, the role 
of Cu-centered NPLEs shall not be ruled out, since the associated 
PBs are weaker; with mean -ICOHP values distributed around about 0.8 eV. 

Within this context, it is possible to show that the computed mean -ICOHP 
values are also correlated with the nature of the \textit{bond exchange} 
processes; and propose that such processes in Zr-centered NPLEs are more 
likely to be associated to local rearrangements with low mobility -- just 
as depicted in Fig. S37 for the example Zr-centered NPLE. On the other hand, 
\textit{bond exchange} processes going on in Cu-centered NPLEs could be 
said more likely to be related to a higher mobility. Without claiming to 
make quantitative predictions on the rheology of these MGs, the mean-squared 
displacements perpendicular to the loading direction (MSD$_{YZ}$) averaged 
over atoms of the same specie were also computed. These bare results are 
shown in Figs. S40 and S41, and they can be used to gain a qualitative view 
of atomic mobility upon uniaxial compression up to 5\% strain. In short, 
it can be seen that the overall atomic mobility in the MG 
Zr$_{49}$Cu$_{49}$Al$_{2}$ is higher. However, in both NCs, Zr and Al atoms 
present equivalent mobility, whereas Cu atoms present greater and discernible 
mobility; especially in that lowest Al-content MG.

Finally, with respect to the NPLEs-BBs, one can see in Figs.~\ref{img:histmeans}(c) 
and (f) that they are not very dependent on the chemical specie of the 
central atom, concerning the ranges of the associated mean -ICOHP values; 
which go from 0.0 up to 0.7 eV for Cu-centered NPLEs, and are a little 
more extended up to about 1.0 eV for the Zr-centered NPLEs. An additional 
remark that is valid for all histograms in Fig.~\ref{img:histmeans} is 
that the coordination numbers of Zr atoms are in general higher (about 
18) than in Cu and Al (about 12). So, it is expected that the frequencies 
of broken bonds in Zr-centered LEs will be higher as well. Nevertheless, 
it is clear that the abovementioned ranges of -ICOHP values verified for 
NPLEs-BBs match those pointed in Fig.~\ref{img:ICOHP_ZCA-x} as the weakest 
bonds in the structures of the studied ZCA alloys (all the Cu--Cu interactions 
and a small fraction of the Zr--Zr and Zr-Cu bonds).

A final note, also related to the histograms shown Fig.~\ref{img:histmeans}, 
concerns the use of the arithmetic mean -ICOHP values averaged over 
individual groups of neighbours in each LE. The mean has just been taken 
as an index able to yield a range of -ICOHP values more compatible with 
those computed for individual bonds shown in Fig.~\ref{img:ICOHP_ZCA-x}. 
The corresponding sums (as shown in Figs. S42 to S47) is an alternative 
option that could be used to discuss the outcomes of the CMD simulations 
in terms of chemical bonding, for instance. Moreover, in order to provide 
an insight into the dispersion of the -ICOHP values in each LE used 
to compute the associated arithmetic means, the corresponding standard 
deviations are also available in Figs. S48 to S53.

\section{Summary and Conclusions}

This work has introduced a machine learning (ML)-based approach that 
brings to research in the field of metallic glasses (MGs) a feasible 
solution for the daunting task of extracting quantum chemical information 
from realistic structural models of such highly complex systems -- 
normally containing thousands of atoms. The quantum mechanical bonding 
descriptors are computed with the crystal orbital Hamilton population 
(COHP) method, with the accuracy of first-principles density functional 
theory calculations; and used as a measure of bond strengths. The 
ML model is founded upon a Gaussian process regression framework 
and the smooth overlap of atomic positions (SOAP) descriptor for atomic 
local environments (LEs).

The ML-based approach has proven effective when applied to MGs of a 
prototypical alloy system; providing, firstly, an unprecedented ``big 
picture'' view of bond strengths between atoms in their structures. 
Next, it has been employed under the specific scope of mechanical 
loading; aiming at looking for insights into the 
drastic change in the experimental mechanical behavior upon alloying of 
Al reported in the literature~\cite{KumarOhkubo2007}. Using ordinary 
descriptive statistics, the resulting overview of chemical bond strengths 
revealed a chemical/structural heterogeneity that is quite in line with 
the propensity for \textit{bond exchange} processes verified for different 
types of LEs in the structures of the studied MGs. Additionally, it also 
enabled the assignment of such processes to migration and rearrangement 
mechanisms; based on identified differentiated atomic mobilities.

It is important to point out that bond strength is a key and enabling 
element for the development of methods for short-range order 
identification and atom categorization in MGs. In fact, introducing 
chemical bonding theory into that specific scope is not trivial, but 
it can has the power to bring a complementary chemical sense 
to those already established tools like Voronoi polyhedrons and common 
neighbor analysis (CNA)~\cite{YueInoue2017}; which are based purely on 
structural topology. Moreover, in addition to bond strengths (-ICOHP 
values), there are other quantum-mechanical indicators of bonding in 
materials that can be learned~\cite{RatySchumacher2019} in the same way. 
Also, the versatility of the SOAP descriptor allows complementarity 
converse approaches to recover detailed information regarding electronic 
structure, which can be interfaced with the proposed ML model without 
much effort. For example, the detailed features (bonding, nonbonding, 
and antibonding) of the full COHP curve of a given LE in a realistic 
structural model used in a classical molecular dynamics simulation, 
can be attained by looking for the best matching LE in the database of 
interactions. 

Regarding the application of the ML model presented in this work, 
although rather restrained -- but quite consistent -- the exposed 
intermix of chemical/structural and dynamical inhomogeneities 
existing in the studied MGs certainly paves the way towards innovative 
approaches from the perspective of chemical bonds. There are promising 
applications in plenty of other contexts in which atom categorization 
based on the strength of chemical bonds plays a key role. Among those, 
it is possible to highlight: the interpretation of dynamical mechanical 
analysis experimental results~\cite{QiaoWang2019}, for unveiling 
atomic-scale mechanisms related to phenomena like viscoelastic behavior 
and internal friction; and also studies of shear bands~\cite{GreerCheng2013} 
for gaining insights into, for instance, the role of the non-persistent 
local environments in the MG Zr$_{49}$Cu$_{49}$Al$_{2}$ on 
nucleation of nanocrystals formed upon deformation, as reported in the 
literature~\cite{KumarOhkubo2007}.

\begin{acknowledgements}
The author is especially thankful to Prof. G\'{a}bor Cs\'{a}nyi from 
University of Cambridge, UK for providing valuable guidelines and 
suggestions on the implementation of the ML model. Many thanks for 
fruitful discussions to Prof. Jichao Qiao from Northwestern Polytechnical 
University, China. The author also acknowledges a fellowship from the 
S\~{a}o Paulo State Research Foundation (FAPESP; grant 2016/12319--0) 
and thanks UFSCar and Prof. Jos\'{e} P. Rino for support. This 
work used the Petaflop computing facilities and associated support 
services of SDumont, provided by the Brazilian National Laboratory for 
Scientific Computing (LNCC) in Petr\'{o}polis, RJ. It also has to be 
pointed that about 20\% of the CMD simulations were carried out using 
the high-performance computing facilities and associated support services 
of the National Center for High Performance Computing in S\~{a}o 
Paulo (CENAPAD-SP); which is also acknowledged.
\end{acknowledgements}

\bibliography{apssamp}

\begin{thebibliography}{56}%
\makeatletter
\providecommand \@ifxundefined [1]{%
 \@ifx{#1\undefined}
}%
\providecommand \@ifnum [1]{%
 \ifnum #1\expandafter \@firstoftwo
 \else \expandafter \@secondoftwo
 \fi
}%
\providecommand \@ifx [1]{%
 \ifx #1\expandafter \@firstoftwo
 \else \expandafter \@secondoftwo
 \fi
}%
\providecommand \natexlab [1]{#1}%
\providecommand \enquote  [1]{``#1''}%
\providecommand \bibnamefont  [1]{#1}%
\providecommand \bibfnamefont [1]{#1}%
\providecommand \citenamefont [1]{#1}%
\providecommand \href@noop [0]{\@secondoftwo}%
\providecommand \href [0]{\begingroup \@sanitize@url \@href}%
\providecommand \@href[1]{\@@startlink{#1}\@@href}%
\providecommand \@@href[1]{\endgroup#1\@@endlink}%
\providecommand \@sanitize@url [0]{\catcode `\\12\catcode `\$12\catcode
  `\&12\catcode `\#12\catcode `\^12\catcode `\_12\catcode `\%12\relax}%
\providecommand \@@startlink[1]{}%
\providecommand \@@endlink[0]{}%
\providecommand \url  [0]{\begingroup\@sanitize@url \@url }%
\providecommand \@url [1]{\endgroup\@href {#1}{\urlprefix }}%
\providecommand \urlprefix  [0]{URL }%
\providecommand \Eprint [0]{\href }%
\providecommand \doibase [0]{http://dx.doi.org/}%
\providecommand \selectlanguage [0]{\@gobble}%
\providecommand \bibinfo  [0]{\@secondoftwo}%
\providecommand \bibfield  [0]{\@secondoftwo}%
\providecommand \translation [1]{[#1]}%
\providecommand \BibitemOpen [0]{}%
\providecommand \bibitemStop [0]{}%
\providecommand \bibitemNoStop [0]{.\EOS\space}%
\providecommand \EOS [0]{\spacefactor3000\relax}%
\providecommand \BibitemShut  [1]{\csname bibitem#1\endcsname}%
\let\auto@bib@innerbib\@empty
\bibitem [{\citenamefont {Klement}\ \emph {et~al.}(1960)\citenamefont
  {Klement}, \citenamefont {Willens},\ and\ \citenamefont
  {Duwez}}]{KlementWillens1960}%
  \BibitemOpen
  \bibfield  {author} {\bibinfo {author} {\bibfnamefont {W.}~\bibnamefont
  {Klement}}, \bibinfo {author} {\bibfnamefont {R.~H.}\ \bibnamefont
  {Willens}}, \ and\ \bibinfo {author} {\bibfnamefont {P.}~\bibnamefont
  {Duwez}},\ }\href {\doibase 10.1038/187869b0} {\bibfield  {journal} {\bibinfo
   {journal} {Nature}\ }\textbf {\bibinfo {volume} {187}},\ \bibinfo {pages}
  {869} (\bibinfo {year} {1960})}\BibitemShut {NoStop}%
\bibitem [{\citenamefont {Milanez}\ \emph {et~al.}(2017)\citenamefont
  {Milanez}, \citenamefont {Faria}, \citenamefont {Leiva}, \citenamefont
  {Kiminami},\ and\ \citenamefont {Botta}}]{MilanezFaria2017}%
  \BibitemOpen
  \bibfield  {author} {\bibinfo {author} {\bibfnamefont {D.~H.}\ \bibnamefont
  {Milanez}}, \bibinfo {author} {\bibfnamefont {L.~I.~L.}\ \bibnamefont
  {Faria}}, \bibinfo {author} {\bibfnamefont {D.~R.}\ \bibnamefont {Leiva}},
  \bibinfo {author} {\bibfnamefont {C.~S.}\ \bibnamefont {Kiminami}}, \ and\
  \bibinfo {author} {\bibfnamefont {W.~J.}\ \bibnamefont {Botta}},\ }\href
  {\doibase 10.1016/j.jallcom.2017.05.105} {\bibfield  {journal} {\bibinfo
  {journal} {J. Alloys Compd.}\ }\textbf {\bibinfo {volume} {716}},\ \bibinfo
  {pages} {330} (\bibinfo {year} {2017})}\BibitemShut {NoStop}%
\bibitem [{\citenamefont {Chen}(2011)}]{Chen2011}%
  \BibitemOpen
  \bibfield  {author} {\bibinfo {author} {\bibfnamefont {M.}~\bibnamefont
  {Chen}},\ }\href {\doibase 10.1038/asiamat.2011.30} {\bibfield  {journal}
  {\bibinfo  {journal} {NPG Asia Mater.}\ }\textbf {\bibinfo {volume} {3}},\
  \bibinfo {pages} {82} (\bibinfo {year} {2011})}\BibitemShut {NoStop}%
\bibitem [{\citenamefont {Jiao}\ \emph {et~al.}(2017)\citenamefont {Jiao},
  \citenamefont {Liu}, \citenamefont {Lin}, \citenamefont {Zhou}, \citenamefont
  {Wang}, \citenamefont {Fujita}, \citenamefont {Hirata}, \citenamefont {Li},\
  and\ \citenamefont {Chen}}]{JiaoLiu2017}%
  \BibitemOpen
  \bibfield  {author} {\bibinfo {author} {\bibfnamefont {W.}~\bibnamefont
  {Jiao}}, \bibinfo {author} {\bibfnamefont {P.}~\bibnamefont {Liu}}, \bibinfo
  {author} {\bibfnamefont {H.}~\bibnamefont {Lin}}, \bibinfo {author}
  {\bibfnamefont {W.}~\bibnamefont {Zhou}}, \bibinfo {author} {\bibfnamefont
  {Z.}~\bibnamefont {Wang}}, \bibinfo {author} {\bibfnamefont {T.}~\bibnamefont
  {Fujita}}, \bibinfo {author} {\bibfnamefont {A.}~\bibnamefont {Hirata}},
  \bibinfo {author} {\bibfnamefont {H.-W.}\ \bibnamefont {Li}}, \ and\ \bibinfo
  {author} {\bibfnamefont {M.}~\bibnamefont {Chen}},\ }\href {\doibase
  10.1021/acs.chemmater.7b01038} {\bibfield  {journal} {\bibinfo  {journal}
  {Chem. Mat.}\ }\textbf {\bibinfo {volume} {29}},\ \bibinfo {pages} {4478}
  (\bibinfo {year} {2017})}\BibitemShut {NoStop}%
\bibitem [{\citenamefont {Mahbooba}\ \emph {et~al.}(2018)\citenamefont
  {Mahbooba}, \citenamefont {Thorsson}, \citenamefont {Unosson}, \citenamefont
  {Skoglund}, \citenamefont {West}, \citenamefont {Horn}, \citenamefont {Rock},
  \citenamefont {Vogli},\ and\ \citenamefont
  {Harrysson}}]{MahboobaThorsson2018}%
  \BibitemOpen
  \bibfield  {author} {\bibinfo {author} {\bibfnamefont {Z.}~\bibnamefont
  {Mahbooba}}, \bibinfo {author} {\bibfnamefont {L.}~\bibnamefont {Thorsson}},
  \bibinfo {author} {\bibfnamefont {M.}~\bibnamefont {Unosson}}, \bibinfo
  {author} {\bibfnamefont {P.}~\bibnamefont {Skoglund}}, \bibinfo {author}
  {\bibfnamefont {H.}~\bibnamefont {West}}, \bibinfo {author} {\bibfnamefont
  {T.}~\bibnamefont {Horn}}, \bibinfo {author} {\bibfnamefont {C.}~\bibnamefont
  {Rock}}, \bibinfo {author} {\bibfnamefont {E.}~\bibnamefont {Vogli}}, \ and\
  \bibinfo {author} {\bibfnamefont {O.}~\bibnamefont {Harrysson}},\ }\href
  {\doibase 10.1016/j.apmt.2018.02.011} {\bibfield  {journal} {\bibinfo
  {journal} {Appl. Mater. Today}\ }\textbf {\bibinfo {volume} {11}},\ \bibinfo
  {pages} {264} (\bibinfo {year} {2018})}\BibitemShut {NoStop}%
\bibitem [{\citenamefont {Li}\ \emph {et~al.}(2018)\citenamefont {Li},
  \citenamefont {Zhang}, \citenamefont {Xing}, \citenamefont {Ouyang},\ and\
  \citenamefont {Liu}}]{LiZhang2018}%
  \BibitemOpen
  \bibfield  {author} {\bibinfo {author} {\bibfnamefont {N.}~\bibnamefont
  {Li}}, \bibinfo {author} {\bibfnamefont {J.}~\bibnamefont {Zhang}}, \bibinfo
  {author} {\bibfnamefont {W.}~\bibnamefont {Xing}}, \bibinfo {author}
  {\bibfnamefont {D.}~\bibnamefont {Ouyang}}, \ and\ \bibinfo {author}
  {\bibfnamefont {L.}~\bibnamefont {Liu}},\ }\href {\doibase
  10.1016/j.matdes.2018.01.061} {\bibfield  {journal} {\bibinfo  {journal}
  {Mater. \& Design}\ }\textbf {\bibinfo {volume} {143}},\ \bibinfo {pages}
  {285} (\bibinfo {year} {2018})}\BibitemShut {NoStop}%
\bibitem [{\citenamefont {Kruzic}(2016)}]{Kruzic2016}%
  \BibitemOpen
  \bibfield  {author} {\bibinfo {author} {\bibfnamefont {J.~J.}\ \bibnamefont
  {Kruzic}},\ }\href {\doibase 10.1002/adem.201600066} {\bibfield  {journal}
  {\bibinfo  {journal} {Adv. Eng. Mater.}\ }\textbf {\bibinfo {volume} {18}},\
  \bibinfo {pages} {1308} (\bibinfo {year} {2016})}\BibitemShut {NoStop}%
\bibitem [{\citenamefont {Qiao}\ \emph {et~al.}(2019)\citenamefont {Qiao},
  \citenamefont {Wang}, \citenamefont {Pelletier}, \citenamefont {Kato},
  \citenamefont {Casalini}, \citenamefont {Crespo}, \citenamefont {Pineda},
  \citenamefont {Yao},\ and\ \citenamefont {Yang}}]{QiaoWang2019}%
  \BibitemOpen
  \bibfield  {author} {\bibinfo {author} {\bibfnamefont {J.~C.}\ \bibnamefont
  {Qiao}}, \bibinfo {author} {\bibfnamefont {Q.}~\bibnamefont {Wang}}, \bibinfo
  {author} {\bibfnamefont {J.~M.}\ \bibnamefont {Pelletier}}, \bibinfo {author}
  {\bibfnamefont {H.}~\bibnamefont {Kato}}, \bibinfo {author} {\bibfnamefont
  {R.}~\bibnamefont {Casalini}}, \bibinfo {author} {\bibfnamefont
  {D.}~\bibnamefont {Crespo}}, \bibinfo {author} {\bibfnamefont
  {E.}~\bibnamefont {Pineda}}, \bibinfo {author} {\bibfnamefont
  {Y.}~\bibnamefont {Yao}}, \ and\ \bibinfo {author} {\bibfnamefont
  {Y.}~\bibnamefont {Yang}},\ }\href {\doibase 10.1016/j.pmatsci.2019.04.005}
  {\bibfield  {journal} {\bibinfo  {journal} {Prog. Mater. Sci.}\ }\textbf
  {\bibinfo {volume} {104}},\ \bibinfo {pages} {250} (\bibinfo {year}
  {2019})}\BibitemShut {NoStop}%
\bibitem [{\citenamefont {Wang}\ and\ \citenamefont
  {Wang}(2018)}]{WangWang2018}%
  \BibitemOpen
  \bibfield  {author} {\bibinfo {author} {\bibfnamefont {Z.}~\bibnamefont
  {Wang}}\ and\ \bibinfo {author} {\bibfnamefont {W.-H.}\ \bibnamefont
  {Wang}},\ }\href {\doibase 10.1093/nsr/nwy084} {\bibfield  {journal}
  {\bibinfo  {journal} {National Sci. Rev.}\ }\textbf {\bibinfo {volume} {6}},\
  \bibinfo {pages} {304} (\bibinfo {year} {2018})}\BibitemShut {NoStop}%
\bibitem [{\citenamefont {Yu}\ \emph {et~al.}(2013)\citenamefont {Yu},
  \citenamefont {Samwer}, \citenamefont {Wang},\ and\ \citenamefont
  {Bai}}]{YuSamwer2013}%
  \BibitemOpen
  \bibfield  {author} {\bibinfo {author} {\bibfnamefont {H.~B.}\ \bibnamefont
  {Yu}}, \bibinfo {author} {\bibfnamefont {K.}~\bibnamefont {Samwer}}, \bibinfo
  {author} {\bibfnamefont {W.~H.}\ \bibnamefont {Wang}}, \ and\ \bibinfo
  {author} {\bibfnamefont {H.~Y.}\ \bibnamefont {Bai}},\ }\href {\doibase
  10.1038/ncomms3204} {\bibfield  {journal} {\bibinfo  {journal} {Nature
  Comm.}\ }\textbf {\bibinfo {volume} {4}},\ \bibinfo {pages} {2204} (\bibinfo
  {year} {2013})}\BibitemShut {NoStop}%
\bibitem [{\citenamefont {Yu}\ \emph {et~al.}(2014)\citenamefont {Yu},
  \citenamefont {Wang}, \citenamefont {Bai},\ and\ \citenamefont
  {Samwer}}]{YuWang2014}%
  \BibitemOpen
  \bibfield  {author} {\bibinfo {author} {\bibfnamefont {H.~B.}\ \bibnamefont
  {Yu}}, \bibinfo {author} {\bibfnamefont {W.-H.}\ \bibnamefont {Wang}},
  \bibinfo {author} {\bibfnamefont {H.~Y.}\ \bibnamefont {Bai}}, \ and\
  \bibinfo {author} {\bibfnamefont {K.}~\bibnamefont {Samwer}},\ }\href
  {\doibase 10.1093/nsr/nwu018} {\bibfield  {journal} {\bibinfo  {journal}
  {National Sci. Rev.}\ }\textbf {\bibinfo {volume} {1}},\ \bibinfo {pages}
  {429} (\bibinfo {year} {2014})}\BibitemShut {NoStop}%
\bibitem [{\citenamefont {Qiao}\ and\ \citenamefont
  {Pelletier}(2014)}]{QiaoPelletier2014}%
  \BibitemOpen
  \bibfield  {author} {\bibinfo {author} {\bibfnamefont {J.~C.}\ \bibnamefont
  {Qiao}}\ and\ \bibinfo {author} {\bibfnamefont {J.~M.}\ \bibnamefont
  {Pelletier}},\ }\href {\doibase 10.1016/j.jmst.2014.04.018} {\bibfield
  {journal} {\bibinfo  {journal} {J. Mater. Sci. Technol.}\ }\textbf {\bibinfo
  {volume} {30}},\ \bibinfo {pages} {523} (\bibinfo {year} {2014})}\BibitemShut
  {NoStop}%
\bibitem [{\citenamefont {Qiao}\ \emph {et~al.}(2017)\citenamefont {Qiao},
  \citenamefont {Wang}, \citenamefont {Crespo}, \citenamefont {Yang},\ and\
  \citenamefont {Pelletier}}]{QiaoWang2017}%
  \BibitemOpen
  \bibfield  {author} {\bibinfo {author} {\bibfnamefont {J.~C.}\ \bibnamefont
  {Qiao}}, \bibinfo {author} {\bibfnamefont {Q.}~\bibnamefont {Wang}}, \bibinfo
  {author} {\bibfnamefont {D.}~\bibnamefont {Crespo}}, \bibinfo {author}
  {\bibfnamefont {Y.}~\bibnamefont {Yang}}, \ and\ \bibinfo {author}
  {\bibfnamefont {J.~M.}\ \bibnamefont {Pelletier}},\ }\href {\doibase
  10.1088/1674-1056/26/1/016402} {\bibfield  {journal} {\bibinfo  {journal}
  {Chin. Phys. B}\ }\textbf {\bibinfo {volume} {26}},\ \bibinfo {pages}
  {016402} (\bibinfo {year} {2017})}\BibitemShut {NoStop}%
\bibitem [{\citenamefont {Yu}\ \emph {et~al.}(2017)\citenamefont {Yu},
  \citenamefont {Richert},\ and\ \citenamefont {Samwer}}]{YuRichert2017}%
  \BibitemOpen
  \bibfield  {author} {\bibinfo {author} {\bibfnamefont {H.~B.}\ \bibnamefont
  {Yu}}, \bibinfo {author} {\bibfnamefont {R.}~\bibnamefont {Richert}}, \ and\
  \bibinfo {author} {\bibfnamefont {K.}~\bibnamefont {Samwer}},\ }\href
  {\doibase 10.1126/sciadv.1701577} {\bibfield  {journal} {\bibinfo  {journal}
  {Sci. Adv.}\ }\textbf {\bibinfo {volume} {3}},\ \bibinfo {pages} {e1701577}
  (\bibinfo {year} {2017})}\BibitemShut {NoStop}%
\bibitem [{\citenamefont {Cheng}\ \emph {et~al.}(2009)\citenamefont {Cheng},
  \citenamefont {Ma},\ and\ \citenamefont {Sheng}}]{ChengMa2009}%
  \BibitemOpen
  \bibfield  {author} {\bibinfo {author} {\bibfnamefont {Y.~Q.}\ \bibnamefont
  {Cheng}}, \bibinfo {author} {\bibfnamefont {E.}~\bibnamefont {Ma}}, \ and\
  \bibinfo {author} {\bibfnamefont {H.~W.}\ \bibnamefont {Sheng}},\ }\href
  {\doibase 10.1103/PhysRevLett.102.245501} {\bibfield  {journal} {\bibinfo
  {journal} {Phys. Rev. Lett.}\ }\textbf {\bibinfo {volume} {102}},\ \bibinfo
  {pages} {245501} (\bibinfo {year} {2009})}\BibitemShut {NoStop}%
\bibitem [{\citenamefont {Bart\'{o}k}\ \emph {et~al.}(2017)\citenamefont
  {Bart\'{o}k}, \citenamefont {De}, \citenamefont {Poelking}, \citenamefont
  {Bernstein}, \citenamefont {Kermode}, \citenamefont {Cs\'{a}nyi},\ and\
  \citenamefont {Ceriotti}}]{BartokDe2017}%
  \BibitemOpen
  \bibfield  {author} {\bibinfo {author} {\bibfnamefont {A.~P.}\ \bibnamefont
  {Bart\'{o}k}}, \bibinfo {author} {\bibfnamefont {S.}~\bibnamefont {De}},
  \bibinfo {author} {\bibfnamefont {C.}~\bibnamefont {Poelking}}, \bibinfo
  {author} {\bibfnamefont {N.}~\bibnamefont {Bernstein}}, \bibinfo {author}
  {\bibfnamefont {J.~R.}\ \bibnamefont {Kermode}}, \bibinfo {author}
  {\bibfnamefont {G.}~\bibnamefont {Cs\'{a}nyi}}, \ and\ \bibinfo {author}
  {\bibfnamefont {M.}~\bibnamefont {Ceriotti}},\ }\href {\doibase
  10.1126/sciadv.1701816} {\bibfield  {journal} {\bibinfo  {journal} {Science
  Advances}\ }\textbf {\bibinfo {volume} {3}},\ \bibinfo {pages} {e1701816}
  (\bibinfo {year} {2017})}\BibitemShut {NoStop}%
\bibitem [{\citenamefont {Deringer}\ \emph {et~al.}(2014)\citenamefont
  {Deringer}, \citenamefont {Zhang}, \citenamefont {Lumeij}, \citenamefont
  {Maintz}, \citenamefont {Wuttig}, \citenamefont {Mazzarello},\ and\
  \citenamefont {Dronskowski}}]{DeringerZhang2014}%
  \BibitemOpen
  \bibfield  {author} {\bibinfo {author} {\bibfnamefont {V.~L.}\ \bibnamefont
  {Deringer}}, \bibinfo {author} {\bibfnamefont {W.}~\bibnamefont {Zhang}},
  \bibinfo {author} {\bibfnamefont {M.}~\bibnamefont {Lumeij}}, \bibinfo
  {author} {\bibfnamefont {S.}~\bibnamefont {Maintz}}, \bibinfo {author}
  {\bibfnamefont {M.}~\bibnamefont {Wuttig}}, \bibinfo {author} {\bibfnamefont
  {R.}~\bibnamefont {Mazzarello}}, \ and\ \bibinfo {author} {\bibfnamefont
  {R.}~\bibnamefont {Dronskowski}},\ }\href {\doibase 10.1002/anie.201404223}
  {\bibfield  {journal} {\bibinfo  {journal} {Angew. Chem. Int. Ed.}\ }\textbf
  {\bibinfo {volume} {53}},\ \bibinfo {pages} {10817} (\bibinfo {year}
  {2014})}\BibitemShut {NoStop}%
\bibitem [{\citenamefont {Hohenberg}\ and\ \citenamefont
  {Kohn}(1964)}]{HohenbergKohn1964}%
  \BibitemOpen
  \bibfield  {author} {\bibinfo {author} {\bibfnamefont {P.}~\bibnamefont
  {Hohenberg}}\ and\ \bibinfo {author} {\bibfnamefont {W.}~\bibnamefont
  {Kohn}},\ }\href {\doibase 10.1103/PhysRev.136.B864} {\bibfield  {journal}
  {\bibinfo  {journal} {Phys. Rev.}\ }\textbf {\bibinfo {volume} {136}},\
  \bibinfo {pages} {B864} (\bibinfo {year} {1964})}\BibitemShut {NoStop}%
\bibitem [{\citenamefont {Kohn}\ and\ \citenamefont
  {Sham}(1965)}]{KohnSham1965}%
  \BibitemOpen
  \bibfield  {author} {\bibinfo {author} {\bibfnamefont {W.}~\bibnamefont
  {Kohn}}\ and\ \bibinfo {author} {\bibfnamefont {L.~J.}\ \bibnamefont
  {Sham}},\ }\href {\doibase 10.1103/PhysRev.140.A1133} {\bibfield  {journal}
  {\bibinfo  {journal} {Phys. Rev.}\ }\textbf {\bibinfo {volume} {140}},\
  \bibinfo {pages} {A1133} (\bibinfo {year} {1965})}\BibitemShut {NoStop}%
\bibitem [{\citenamefont {Dronskowski}\ and\ \citenamefont
  {Bl\"{o}chl}(1993)}]{DronskowskiBlochl1993}%
  \BibitemOpen
  \bibfield  {author} {\bibinfo {author} {\bibfnamefont {R.}~\bibnamefont
  {Dronskowski}}\ and\ \bibinfo {author} {\bibfnamefont {P.~E.}\ \bibnamefont
  {Bl\"{o}chl}},\ }\href {\doibase 10.1021/j100135a014} {\bibfield  {journal}
  {\bibinfo  {journal} {J. Phys. Chem.}\ }\textbf {\bibinfo {volume} {97}},\
  \bibinfo {pages} {8617} (\bibinfo {year} {1993})}\BibitemShut {NoStop}%
\bibitem [{\citenamefont {Pauly}\ \emph {et~al.}(2010)\citenamefont {Pauly},
  \citenamefont {Liu}, \citenamefont {Gorantla}, \citenamefont {Wang},
  \citenamefont {K\"{u}hn}, \citenamefont {Kim},\ and\ \citenamefont
  {Eckert}}]{PaulyLiu2010}%
  \BibitemOpen
  \bibfield  {author} {\bibinfo {author} {\bibfnamefont {S.}~\bibnamefont
  {Pauly}}, \bibinfo {author} {\bibfnamefont {G.}~\bibnamefont {Liu}}, \bibinfo
  {author} {\bibfnamefont {S.}~\bibnamefont {Gorantla}}, \bibinfo {author}
  {\bibfnamefont {G.}~\bibnamefont {Wang}}, \bibinfo {author} {\bibfnamefont
  {U.}~\bibnamefont {K\"{u}hn}}, \bibinfo {author} {\bibfnamefont {D.~H.}\
  \bibnamefont {Kim}}, \ and\ \bibinfo {author} {\bibfnamefont
  {J.}~\bibnamefont {Eckert}},\ }\href {\doibase 10.1016/j.actamat.2010.05.026}
  {\bibfield  {journal} {\bibinfo  {journal} {Acta Mater.}\ }\textbf {\bibinfo
  {volume} {58}},\ \bibinfo {pages} {4883} (\bibinfo {year}
  {2010})}\BibitemShut {NoStop}%
\bibitem [{\citenamefont {Wang}\ \emph {et~al.}(2018)\citenamefont {Wang},
  \citenamefont {Inoue}, \citenamefont {Zhao}, \citenamefont {Kongc},
  \citenamefont {Zhu}, \citenamefont {Kaban}, \citenamefont {Stoica},
  \citenamefont {Oswald}, \citenamefont {Fan}, \citenamefont {Shalaan},
  \citenamefont {Al-Marzouki}, \citenamefont {Eckert}, \citenamefont {Yin},\
  and\ \citenamefont {Li}}]{WangInoue2018}%
  \BibitemOpen
  \bibfield  {author} {\bibinfo {author} {\bibfnamefont {X.~H.}\ \bibnamefont
  {Wang}}, \bibinfo {author} {\bibfnamefont {A.}~\bibnamefont {Inoue}},
  \bibinfo {author} {\bibfnamefont {J.~F.}\ \bibnamefont {Zhao}}, \bibinfo
  {author} {\bibfnamefont {F.~L.}\ \bibnamefont {Kongc}}, \bibinfo {author}
  {\bibfnamefont {S.~L.}\ \bibnamefont {Zhu}}, \bibinfo {author} {\bibfnamefont
  {I.}~\bibnamefont {Kaban}}, \bibinfo {author} {\bibfnamefont
  {M.}~\bibnamefont {Stoica}}, \bibinfo {author} {\bibfnamefont
  {S.}~\bibnamefont {Oswald}}, \bibinfo {author} {\bibfnamefont
  {C.}~\bibnamefont {Fan}}, \bibinfo {author} {\bibfnamefont {E.}~\bibnamefont
  {Shalaan}}, \bibinfo {author} {\bibfnamefont {F.}~\bibnamefont
  {Al-Marzouki}}, \bibinfo {author} {\bibfnamefont {J.}~\bibnamefont {Eckert}},
  \bibinfo {author} {\bibfnamefont {F.~X.}\ \bibnamefont {Yin}}, \ and\
  \bibinfo {author} {\bibfnamefont {Q.}~\bibnamefont {Li}},\ }\href {\doibase
  10.1016/j.jallcom.2017.12.318} {\bibfield  {journal} {\bibinfo  {journal} {J.
  Alloys Compd.}\ }\textbf {\bibinfo {volume} {739}},\ \bibinfo {pages} {1104}
  (\bibinfo {year} {2018})}\BibitemShut {NoStop}%
\bibitem [{\citenamefont {Pekin}\ \emph {et~al.}(2019)\citenamefont {Pekin},
  \citenamefont {Ding}, \citenamefont {Gammer}, \citenamefont {Ozdol},
  \citenamefont {Ophus}, \citenamefont {Asta}, \citenamefont {Ritchie},\ and\
  \citenamefont {Minor}}]{PekinDing2019}%
  \BibitemOpen
  \bibfield  {author} {\bibinfo {author} {\bibfnamefont {T.~C.}\ \bibnamefont
  {Pekin}}, \bibinfo {author} {\bibfnamefont {J.}~\bibnamefont {Ding}},
  \bibinfo {author} {\bibfnamefont {C.}~\bibnamefont {Gammer}}, \bibinfo
  {author} {\bibfnamefont {B.}~\bibnamefont {Ozdol}}, \bibinfo {author}
  {\bibfnamefont {C.}~\bibnamefont {Ophus}}, \bibinfo {author} {\bibfnamefont
  {M.}~\bibnamefont {Asta}}, \bibinfo {author} {\bibfnamefont {R.~O.}\
  \bibnamefont {Ritchie}}, \ and\ \bibinfo {author} {\bibfnamefont {A.~M.}\
  \bibnamefont {Minor}},\ }\href {\doibase 10.1038/s41467-019-10416-5}
  {\bibfield  {journal} {\bibinfo  {journal} {Nature Mater.}\ }\textbf
  {\bibinfo {volume} {10}},\ \bibinfo {pages} {2445} (\bibinfo {year}
  {2019})}\BibitemShut {NoStop}%
\bibitem [{\citenamefont {Kumar}\ \emph {et~al.}(2007)\citenamefont {Kumar},
  \citenamefont {Ohkubo}, \citenamefont {Mukai},\ and\ \citenamefont
  {Hono}}]{KumarOhkubo2007}%
  \BibitemOpen
  \bibfield  {author} {\bibinfo {author} {\bibfnamefont {G.}~\bibnamefont
  {Kumar}}, \bibinfo {author} {\bibfnamefont {T.}~\bibnamefont {Ohkubo}},
  \bibinfo {author} {\bibfnamefont {T.}~\bibnamefont {Mukai}}, \ and\ \bibinfo
  {author} {\bibfnamefont {K.}~\bibnamefont {Hono}},\ }\href {\doibase
  10.1016/j.scriptamat.2007.02.013} {\bibfield  {journal} {\bibinfo  {journal}
  {Scripta Mater.}\ }\textbf {\bibinfo {volume} {57}},\ \bibinfo {pages} {173}
  (\bibinfo {year} {2007})}\BibitemShut {NoStop}%
\bibitem [{\citenamefont {Mulliken}(1955)}]{Mulliken1955}%
  \BibitemOpen
  \bibfield  {author} {\bibinfo {author} {\bibfnamefont {R.~S.}\ \bibnamefont
  {Mulliken}},\ }\href {\doibase 10.1063/1.1740589} {\bibfield  {journal}
  {\bibinfo  {journal} {J. Chem. Phys.}\ }\textbf {\bibinfo {volume} {23}},\
  \bibinfo {pages} {1841} (\bibinfo {year} {1955})}\BibitemShut {NoStop}%
\bibitem [{\citenamefont {Hughbanks}\ and\ \citenamefont
  {Hoffmann}(1983)}]{HughbanksHoffmann1983}%
  \BibitemOpen
  \bibfield  {author} {\bibinfo {author} {\bibfnamefont {T.}~\bibnamefont
  {Hughbanks}}\ and\ \bibinfo {author} {\bibfnamefont {R.}~\bibnamefont
  {Hoffmann}},\ }\href {\doibase 10.1021/ja00349a027} {\bibfield  {journal}
  {\bibinfo  {journal} {J. Amer. Chem. Soc.}\ }\textbf {\bibinfo {volume}
  {105}},\ \bibinfo {pages} {3528} (\bibinfo {year} {1983})}\BibitemShut
  {NoStop}%
\bibitem [{\citenamefont {\'{L}ubom\'{i}r Benco}(1995)}]{Benco1995}%
  \BibitemOpen
  \bibfield  {author} {\bibinfo {author} {\bibnamefont {\'{L}ubom\'{i}r
  Benco}},\ }\href {\doibase 10.1016/0038-1098(95)00175-1} {\bibfield
  {journal} {\bibinfo  {journal} {Solid State Comm.}\ }\textbf {\bibinfo
  {volume} {94}},\ \bibinfo {pages} {861} (\bibinfo {year} {1995})}\BibitemShut
  {NoStop}%
\bibitem [{\citenamefont {Sheng}\ \emph {et~al.}(2006)\citenamefont {Sheng},
  \citenamefont {Luo}, \citenamefont {Alamgir}, \citenamefont {Bai},\ and\
  \citenamefont {Ma}}]{ShengLuo2006}%
  \BibitemOpen
  \bibfield  {author} {\bibinfo {author} {\bibfnamefont {H.~W.}\ \bibnamefont
  {Sheng}}, \bibinfo {author} {\bibfnamefont {W.~K.}\ \bibnamefont {Luo}},
  \bibinfo {author} {\bibfnamefont {F.~M.}\ \bibnamefont {Alamgir}}, \bibinfo
  {author} {\bibfnamefont {J.~M.}\ \bibnamefont {Bai}}, \ and\ \bibinfo
  {author} {\bibfnamefont {E.}~\bibnamefont {Ma}},\ }\href {\doibase
  10.1038/nature04421} {\bibfield  {journal} {\bibinfo  {journal} {Nature}\
  }\textbf {\bibinfo {volume} {439}},\ \bibinfo {pages} {419} (\bibinfo {year}
  {2006})}\BibitemShut {NoStop}%
\bibitem [{\citenamefont {Yuan}\ \emph {et~al.}(2019)\citenamefont {Yuan},
  \citenamefont {Yang}, \citenamefont {Xi}, \citenamefont {Shi}, \citenamefont
  {Holland-Moritz}, \citenamefont {Li}, \citenamefont {Hu}, \citenamefont
  {Shen}, \citenamefont {Wang}, \citenamefont {Meyer},\ and\ \citenamefont
  {Wang}}]{YuanYang2019}%
  \BibitemOpen
  \bibfield  {author} {\bibinfo {author} {\bibfnamefont {C.~C.}\ \bibnamefont
  {Yuan}}, \bibinfo {author} {\bibfnamefont {F.}~\bibnamefont {Yang}}, \bibinfo
  {author} {\bibfnamefont {X.~K.}\ \bibnamefont {Xi}}, \bibinfo {author}
  {\bibfnamefont {C.~L.}\ \bibnamefont {Shi}}, \bibinfo {author} {\bibfnamefont
  {D.}~\bibnamefont {Holland-Moritz}}, \bibinfo {author} {\bibfnamefont
  {M.~Z.}\ \bibnamefont {Li}}, \bibinfo {author} {\bibfnamefont
  {F.}~\bibnamefont {Hu}}, \bibinfo {author} {\bibfnamefont {B.~L.}\
  \bibnamefont {Shen}}, \bibinfo {author} {\bibfnamefont {X.~L.}\ \bibnamefont
  {Wang}}, \bibinfo {author} {\bibfnamefont {A.}~\bibnamefont {Meyer}}, \ and\
  \bibinfo {author} {\bibfnamefont {W.~H.}\ \bibnamefont {Wang}},\ }\href
  {\doibase 10.1016/j.mattod.2019.06.001} {\bibfield  {journal} {\bibinfo
  {journal} {Mater. Today}\ } (\bibinfo {year} {2019}),\
  10.1016/j.mattod.2019.06.001}\BibitemShut {NoStop}%
\bibitem [{\citenamefont {Bart\'{o}k}\ \emph {et~al.}(2013)\citenamefont
  {Bart\'{o}k}, \citenamefont {Kondor},\ and\ \citenamefont
  {Cs\'{a}nyi}}]{BartokKondor2013}%
  \BibitemOpen
  \bibfield  {author} {\bibinfo {author} {\bibfnamefont {A.~P.}\ \bibnamefont
  {Bart\'{o}k}}, \bibinfo {author} {\bibfnamefont {R.}~\bibnamefont {Kondor}},
  \ and\ \bibinfo {author} {\bibfnamefont {G.}~\bibnamefont {Cs\'{a}nyi}},\
  }\href {\doibase 10.1103/PhysRevB.87.184115} {\bibfield  {journal} {\bibinfo
  {journal} {Phys. Rev. B}\ }\textbf {\bibinfo {volume} {87}},\ \bibinfo
  {pages} {184115} (\bibinfo {year} {2013})}\BibitemShut {NoStop}%
\bibitem [{\citenamefont {De}\ \emph {et~al.}(2016)\citenamefont {De},
  \citenamefont {Bart\'{o}k}, \citenamefont {Cs\'{a}nyi},\ and\ \citenamefont
  {Ceriotti}}]{DeBartok2016}%
  \BibitemOpen
  \bibfield  {author} {\bibinfo {author} {\bibfnamefont {S.}~\bibnamefont
  {De}}, \bibinfo {author} {\bibfnamefont {A.~P.}\ \bibnamefont {Bart\'{o}k}},
  \bibinfo {author} {\bibfnamefont {G.}~\bibnamefont {Cs\'{a}nyi}}, \ and\
  \bibinfo {author} {\bibfnamefont {M.}~\bibnamefont {Ceriotti}},\ }\href
  {\doibase 10.1039/c6cp00415f} {\bibfield  {journal} {\bibinfo  {journal}
  {Phys. Chem. Chem. Phys.}\ }\textbf {\bibinfo {volume} {18}},\ \bibinfo
  {pages} {13754} (\bibinfo {year} {2016})}\BibitemShut {NoStop}%
\bibitem [{\citenamefont {Plimpton}(1995)}]{Plimpton1995}%
  \BibitemOpen
  \bibfield  {author} {\bibinfo {author} {\bibfnamefont {S.}~\bibnamefont
  {Plimpton}},\ }\href {\doibase 10.1006/jcph.1995.1039} {\bibfield  {journal}
  {\bibinfo  {journal} {J. Comp. Phys.}\ }\textbf {\bibinfo {volume} {117}},\
  \bibinfo {pages} {1} (\bibinfo {year} {1995})}\BibitemShut {NoStop}%
\bibitem [{She()}]{Sheng2011}%
  \BibitemOpen
  \href@noop {} {\enquote {\bibinfo {title} {Eam potentials},}\ }\bibinfo
  {howpublished}
  {\url{https://sites.google.com/site/eampotentials/Home/ZrCuAl}},\ \bibinfo
  {note} {accessed: 2020-07-21}\BibitemShut {NoStop}%
\bibitem [{\citenamefont {Giannozzi}\ \emph {et~al.}(2009)\citenamefont
  {Giannozzi} \emph {et~al.}}]{GiannozziBaroni2009}%
  \BibitemOpen
  \bibfield  {author} {\bibinfo {author} {\bibfnamefont {P.}~\bibnamefont
  {Giannozzi}} \emph {et~al.},\ }\href {\doibase
  10.1088/0953-8984/21/39/395502} {\bibfield  {journal} {\bibinfo  {journal}
  {J. Phys.: Cond. Matter}\ }\textbf {\bibinfo {volume} {21}},\ \bibinfo
  {pages} {395502} (\bibinfo {year} {2009})}\BibitemShut {NoStop}%
\bibitem [{\citenamefont {Giannozzi}\ \emph {et~al.}(2017)\citenamefont
  {Giannozzi} \emph {et~al.}}]{GiannozziAndreussi2017}%
  \BibitemOpen
  \bibfield  {author} {\bibinfo {author} {\bibfnamefont {P.}~\bibnamefont
  {Giannozzi}} \emph {et~al.},\ }\href {\doibase 10.1088/1361-648X/aa8f79}
  {\bibfield  {journal} {\bibinfo  {journal} {J. Phys.: Cond. Matter}\ }\textbf
  {\bibinfo {volume} {29}},\ \bibinfo {pages} {465901} (\bibinfo {year}
  {2017})}\BibitemShut {NoStop}%
\bibitem [{\citenamefont {Bl\"{o}chl}(1994)}]{Blochl1994}%
  \BibitemOpen
  \bibfield  {author} {\bibinfo {author} {\bibfnamefont {P.~E.}\ \bibnamefont
  {Bl\"{o}chl}},\ }\href {\doibase 10.1103/PhysRevB.50.17953} {\bibfield
  {journal} {\bibinfo  {journal} {Phys. Rev. B}\ }\textbf {\bibinfo {volume}
  {50}},\ \bibinfo {pages} {17953} (\bibinfo {year} {1994})}\BibitemShut
  {NoStop}%
\bibitem [{\citenamefont {Corso}(2014)}]{DalCorso2014}%
  \BibitemOpen
  \bibfield  {author} {\bibinfo {author} {\bibfnamefont {A.~D.}\ \bibnamefont
  {Corso}},\ }\href {\doibase 10.1016/j.commatsci.2014.07.043} {\bibfield
  {journal} {\bibinfo  {journal} {Comp. Material Science}\ }\textbf {\bibinfo
  {volume} {95}},\ \bibinfo {pages} {337} (\bibinfo {year} {2014})}\BibitemShut
  {NoStop}%
\bibitem [{\citenamefont {Perdew}\ \emph {et~al.}(1996)\citenamefont {Perdew},
  \citenamefont {Burke},\ and\ \citenamefont {Ernzerhof}}]{PerdewBurke1996}%
  \BibitemOpen
  \bibfield  {author} {\bibinfo {author} {\bibfnamefont {J.~P.}\ \bibnamefont
  {Perdew}}, \bibinfo {author} {\bibfnamefont {K.}~\bibnamefont {Burke}}, \
  and\ \bibinfo {author} {\bibfnamefont {M.}~\bibnamefont {Ernzerhof}},\ }\href
  {\doibase 10.1103/PhysRevLett.77.3865} {\bibfield  {journal} {\bibinfo
  {journal} {Phys. Rev. Lett.}\ }\textbf {\bibinfo {volume} {77}},\ \bibinfo
  {pages} {3865} (\bibinfo {year} {1996})}\BibitemShut {NoStop}%
\bibitem [{\citenamefont {Monkhorst}\ and\ \citenamefont
  {Pack}(1976)}]{MonkhorstPack1976}%
  \BibitemOpen
  \bibfield  {author} {\bibinfo {author} {\bibfnamefont {H.~J.}\ \bibnamefont
  {Monkhorst}}\ and\ \bibinfo {author} {\bibfnamefont {J.~D.}\ \bibnamefont
  {Pack}},\ }\href {\doibase 10.1103/PhysRevB.13.5188} {\bibfield  {journal}
  {\bibinfo  {journal} {Phys. Rev. B}\ }\textbf {\bibinfo {volume} {13}},\
  \bibinfo {pages} {5188} (\bibinfo {year} {1976})}\BibitemShut {NoStop}%
\bibitem [{\citenamefont {Deringer}\ \emph {et~al.}(2011)\citenamefont
  {Deringer}, \citenamefont {Tchougr\'{e}eff},\ and\ \citenamefont
  {Dronskowski}}]{DeringerTchougreeff2011}%
  \BibitemOpen
  \bibfield  {author} {\bibinfo {author} {\bibfnamefont {V.~L.}\ \bibnamefont
  {Deringer}}, \bibinfo {author} {\bibfnamefont {A.~L.}\ \bibnamefont
  {Tchougr\'{e}eff}}, \ and\ \bibinfo {author} {\bibfnamefont {R.}~\bibnamefont
  {Dronskowski}},\ }\href {\doibase 10.1021/jp202489s} {\bibfield  {journal}
  {\bibinfo  {journal} {J. Phys. Chem. A}\ }\textbf {\bibinfo {volume} {115}},\
  \bibinfo {pages} {5461} (\bibinfo {year} {2011})}\BibitemShut {NoStop}%
\bibitem [{\citenamefont {Maintz}\ \emph {et~al.}(2013)\citenamefont {Maintz},
  \citenamefont {Deringer}, \citenamefont {Tchougr\'{e}eff},\ and\
  \citenamefont {Dronskowski}}]{MaintzDeringer2013}%
  \BibitemOpen
  \bibfield  {author} {\bibinfo {author} {\bibfnamefont {S.}~\bibnamefont
  {Maintz}}, \bibinfo {author} {\bibfnamefont {V.~L.}\ \bibnamefont
  {Deringer}}, \bibinfo {author} {\bibfnamefont {A.~L.}\ \bibnamefont
  {Tchougr\'{e}eff}}, \ and\ \bibinfo {author} {\bibfnamefont {R.}~\bibnamefont
  {Dronskowski}},\ }\href {\doibase 10.1002/jcc.23424} {\bibfield  {journal}
  {\bibinfo  {journal} {J. Comp. Chem.}\ }\textbf {\bibinfo {volume} {34}},\
  \bibinfo {pages} {2557} (\bibinfo {year} {2013})}\BibitemShut {NoStop}%
\bibitem [{\citenamefont {Maintz}\ \emph
  {et~al.}(2016{\natexlab{a}})\citenamefont {Maintz}, \citenamefont {Deringer},
  \citenamefont {Tchougr\'{e}eff},\ and\ \citenamefont
  {Dronskowski}}]{MaintzDeringer2016}%
  \BibitemOpen
  \bibfield  {author} {\bibinfo {author} {\bibfnamefont {S.}~\bibnamefont
  {Maintz}}, \bibinfo {author} {\bibfnamefont {V.~L.}\ \bibnamefont
  {Deringer}}, \bibinfo {author} {\bibfnamefont {A.~L.}\ \bibnamefont
  {Tchougr\'{e}eff}}, \ and\ \bibinfo {author} {\bibfnamefont {R.}~\bibnamefont
  {Dronskowski}},\ }\href {\doibase 10.1002/jcc.24300} {\bibfield  {journal}
  {\bibinfo  {journal} {J. Comp. Chem.}\ }\textbf {\bibinfo {volume} {37}},\
  \bibinfo {pages} {1030} (\bibinfo {year} {2016}{\natexlab{a}})}\BibitemShut
  {NoStop}%
\bibitem [{\citenamefont {Maintz}\ \emph
  {et~al.}(2016{\natexlab{b}})\citenamefont {Maintz}, \citenamefont {Esser},\
  and\ \citenamefont {Dronskowski}}]{MaintzEsser2016}%
  \BibitemOpen
  \bibfield  {author} {\bibinfo {author} {\bibfnamefont {S.}~\bibnamefont
  {Maintz}}, \bibinfo {author} {\bibfnamefont {M.}~\bibnamefont {Esser}}, \
  and\ \bibinfo {author} {\bibfnamefont {R.}~\bibnamefont {Dronskowski}},\
  }\href {\doibase 10.5506/APhysPolB.47.1165} {\bibfield  {journal} {\bibinfo
  {journal} {Acta Phys. Pol. B}\ }\textbf {\bibinfo {volume} {47}},\ \bibinfo
  {pages} {1165} (\bibinfo {year} {2016}{\natexlab{b}})}\BibitemShut {NoStop}%
\bibitem [{\citenamefont {Nelson}\ \emph {et~al.}(2017)\citenamefont {Nelson},
  \citenamefont {Konze},\ and\ \citenamefont {Dronskowski}}]{NelsonKonze2017}%
  \BibitemOpen
  \bibfield  {author} {\bibinfo {author} {\bibfnamefont {R.}~\bibnamefont
  {Nelson}}, \bibinfo {author} {\bibfnamefont {P.~M.}\ \bibnamefont {Konze}}, \
  and\ \bibinfo {author} {\bibfnamefont {R.}~\bibnamefont {Dronskowski}},\
  }\href {\doibase 10.1021/acs.jpca.7b08218} {\bibfield  {journal} {\bibinfo
  {journal} {J. Phys. Chem. A}\ }\textbf {\bibinfo {volume} {121}},\ \bibinfo
  {pages} {7778} (\bibinfo {year} {2017})}\BibitemShut {NoStop}%
\bibitem [{\citenamefont {Bunge}\ \emph {et~al.}(1993)\citenamefont {Bunge},
  \citenamefont {Barrientos},\ and\ \citenamefont
  {Bunge}}]{BungeBarrientos1993}%
  \BibitemOpen
  \bibfield  {author} {\bibinfo {author} {\bibfnamefont {C.~F.}\ \bibnamefont
  {Bunge}}, \bibinfo {author} {\bibfnamefont {J.~A.}\ \bibnamefont
  {Barrientos}}, \ and\ \bibinfo {author} {\bibfnamefont {A.}~\bibnamefont
  {Bunge}},\ }\href {\doibase 10.1006/adnd.1993.1003} {\bibfield  {journal}
  {\bibinfo  {journal} {Data Nucl. Data Tables}\ }\textbf {\bibinfo {volume}
  {53}},\ \bibinfo {pages} {113} (\bibinfo {year} {1993})}\BibitemShut
  {NoStop}%
\bibitem [{Gab()}]{Gabor2019}%
  \BibitemOpen
  \href@noop {} {\enquote {\bibinfo {title} {libatoms/quip molecular dynamics
  framework},}\ }\bibinfo {howpublished} {\url{http://www.libatoms.org}},\
  \bibinfo {note} {accessed: 2020-07-21}\BibitemShut {NoStop}%
\bibitem [{\citenamefont {Fan}\ \emph {et~al.}(2009)\citenamefont {Fan},
  \citenamefont {Liaw},\ and\ \citenamefont {Liu}}]{FanLiaw2009}%
  \BibitemOpen
  \bibfield  {author} {\bibinfo {author} {\bibfnamefont {C.}~\bibnamefont
  {Fan}}, \bibinfo {author} {\bibfnamefont {P.~K.}\ \bibnamefont {Liaw}}, \
  and\ \bibinfo {author} {\bibfnamefont {C.~T.}\ \bibnamefont {Liu}},\ }\href
  {\doibase 10.1016/j.intermet.2008.09.007} {\bibfield  {journal} {\bibinfo
  {journal} {Intermetallics}\ }\textbf {\bibinfo {volume} {17}},\ \bibinfo
  {pages} {86} (\bibinfo {year} {2009})}\BibitemShut {NoStop}%
\bibitem [{\citenamefont {Li}\ \emph {et~al.}(2017)\citenamefont {Li},
  \citenamefont {Jiang}, \citenamefont {Ding}, \citenamefont {Peng},
  \citenamefont {Jiang}, \citenamefont {He},\ and\ \citenamefont
  {Sun}}]{LiJiang2017}%
  \BibitemOpen
  \bibfield  {author} {\bibinfo {author} {\bibfnamefont {M.~C.}\ \bibnamefont
  {Li}}, \bibinfo {author} {\bibfnamefont {M.~Q.}\ \bibnamefont {Jiang}},
  \bibinfo {author} {\bibfnamefont {G.}~\bibnamefont {Ding}}, \bibinfo {author}
  {\bibfnamefont {Z.~H.}\ \bibnamefont {Peng}}, \bibinfo {author}
  {\bibfnamefont {F.}~\bibnamefont {Jiang}}, \bibinfo {author} {\bibfnamefont
  {L.}~\bibnamefont {He}}, \ and\ \bibinfo {author} {\bibfnamefont
  {J.}~\bibnamefont {Sun}},\ }\href {\doibase 10.1016/j.jnoncrysol.2017.04.021}
  {\bibfield  {journal} {\bibinfo  {journal} {J. Non-Cryst. Solids}\ }\textbf
  {\bibinfo {volume} {468}},\ \bibinfo {pages} {52} (\bibinfo {year}
  {2017})}\BibitemShut {NoStop}%
\bibitem [{\citenamefont {Jiao}\ \emph {et~al.}(2015)\citenamefont {Jiao},
  \citenamefont {Wang}, \citenamefont {Lan}, \citenamefont {Pan},\ and\
  \citenamefont {Lu}}]{JiaoWang2015}%
  \BibitemOpen
  \bibfield  {author} {\bibinfo {author} {\bibfnamefont {W.}~\bibnamefont
  {Jiao}}, \bibinfo {author} {\bibfnamefont {X.~L.}\ \bibnamefont {Wang}},
  \bibinfo {author} {\bibfnamefont {S.}~\bibnamefont {Lan}}, \bibinfo {author}
  {\bibfnamefont {S.~P.}\ \bibnamefont {Pan}}, \ and\ \bibinfo {author}
  {\bibfnamefont {Z.~P.}\ \bibnamefont {Lu}},\ }\href {\doibase
  10.1063/1.4908122} {\bibfield  {journal} {\bibinfo  {journal} {Appl. Phys.
  Lett.}\ }\textbf {\bibinfo {volume} {106}} (\bibinfo {year} {2015}),\
  10.1063/1.4908122}\BibitemShut {NoStop}%
\bibitem [{\citenamefont {Lekka}\ and\ \citenamefont
  {Evangelakis}(2009)}]{LekkaEvangelakis2009}%
  \BibitemOpen
  \bibfield  {author} {\bibinfo {author} {\bibfnamefont {C.~E.}\ \bibnamefont
  {Lekka}}\ and\ \bibinfo {author} {\bibfnamefont {G.~A.}\ \bibnamefont
  {Evangelakis}},\ }\href {\doibase 10.1016/j.scriptamat.2009.08.008}
  {\bibfield  {journal} {\bibinfo  {journal} {Scripta Mater.}\ }\textbf
  {\bibinfo {volume} {61}},\ \bibinfo {pages} {974} (\bibinfo {year}
  {2009})}\BibitemShut {NoStop}%
\bibitem [{\citenamefont {Lekka}\ \emph {et~al.}(2012)\citenamefont {Lekka},
  \citenamefont {Bokas}, \citenamefont {Almyras}, \citenamefont
  {Papageorgiou},\ and\ \citenamefont {Evangelakis}}]{LekkaBokas2012}%
  \BibitemOpen
  \bibfield  {author} {\bibinfo {author} {\bibfnamefont {C.~E.}\ \bibnamefont
  {Lekka}}, \bibinfo {author} {\bibfnamefont {G.}~\bibnamefont {Bokas}},
  \bibinfo {author} {\bibfnamefont {G.~A.}\ \bibnamefont {Almyras}}, \bibinfo
  {author} {\bibfnamefont {D.~G.}\ \bibnamefont {Papageorgiou}}, \ and\
  \bibinfo {author} {\bibfnamefont {G.~A.}\ \bibnamefont {Evangelakis}},\
  }\href {\doibase 10.1016/j.jallcom.2011.11.038} {\bibfield  {journal}
  {\bibinfo  {journal} {J. Alloys Compd.}\ }\textbf {\bibinfo {volume} {536}},\
  \bibinfo {pages} {S65} (\bibinfo {year} {2012})}\BibitemShut {NoStop}%
\bibitem [{\citenamefont {Melchionna}\ \emph {et~al.}(1993)\citenamefont
  {Melchionna}, \citenamefont {Ciccotti},\ and\ \citenamefont
  {Holian}}]{MelchionnaCiccotti1993}%
  \BibitemOpen
  \bibfield  {author} {\bibinfo {author} {\bibfnamefont {S.}~\bibnamefont
  {Melchionna}}, \bibinfo {author} {\bibfnamefont {G.}~\bibnamefont
  {Ciccotti}}, \ and\ \bibinfo {author} {\bibfnamefont {B.~L.}\ \bibnamefont
  {Holian}},\ }\href {\doibase 10.1080/00268979300100371} {\bibfield  {journal}
  {\bibinfo  {journal} {Mol. Phys.}\ }\textbf {\bibinfo {volume} {78}},\
  \bibinfo {pages} {533} (\bibinfo {year} {1993})}\BibitemShut {NoStop}%
\bibitem [{\citenamefont {Tschopp}\ and\ \citenamefont
  {McDowell}(2008)}]{TschoppMcDowell2008}%
  \BibitemOpen
  \bibfield  {author} {\bibinfo {author} {\bibfnamefont {M.~A.}\ \bibnamefont
  {Tschopp}}\ and\ \bibinfo {author} {\bibfnamefont {D.~L.}\ \bibnamefont
  {McDowell}},\ }\href {\doibase 10.1016/j.jmps.2007.11.012} {\bibfield
  {journal} {\bibinfo  {journal} {J. Mech. Phys. Solids}\ }\textbf {\bibinfo
  {volume} {56}},\ \bibinfo {pages} {1806} (\bibinfo {year}
  {2008})}\BibitemShut {NoStop}%
\bibitem [{\citenamefont {Yue}\ \emph {et~al.}(2017)\citenamefont {Yue},
  \citenamefont {Inoue}, \citenamefont {Liu},\ and\ \citenamefont
  {Fan}}]{YueInoue2017}%
  \BibitemOpen
  \bibfield  {author} {\bibinfo {author} {\bibfnamefont {X.}~\bibnamefont
  {Yue}}, \bibinfo {author} {\bibfnamefont {A.}~\bibnamefont {Inoue}}, \bibinfo
  {author} {\bibfnamefont {C.-T.}\ \bibnamefont {Liu}}, \ and\ \bibinfo
  {author} {\bibfnamefont {C.}~\bibnamefont {Fan}},\ }\href {\doibase
  10.1590/1980-5373-MR-2016-0318} {\bibfield  {journal} {\bibinfo  {journal}
  {Materials Research}\ }\textbf {\bibinfo {volume} {20}},\ \bibinfo {pages}
  {326} (\bibinfo {year} {2017})}\BibitemShut {NoStop}%
\bibitem [{\citenamefont {Raty}\ \emph {et~al.}(2019)\citenamefont {Raty},
  \citenamefont {Schumacher}, \citenamefont {Golub}, \citenamefont {Deringer},
  \citenamefont {Gatti},\ and\ \citenamefont {Wuttig}}]{RatySchumacher2019}%
  \BibitemOpen
  \bibfield  {author} {\bibinfo {author} {\bibfnamefont {J.-Y.}\ \bibnamefont
  {Raty}}, \bibinfo {author} {\bibfnamefont {M.}~\bibnamefont {Schumacher}},
  \bibinfo {author} {\bibfnamefont {P.}~\bibnamefont {Golub}}, \bibinfo
  {author} {\bibfnamefont {V.~L.}\ \bibnamefont {Deringer}}, \bibinfo {author}
  {\bibfnamefont {C.}~\bibnamefont {Gatti}}, \ and\ \bibinfo {author}
  {\bibfnamefont {M.}~\bibnamefont {Wuttig}},\ }\href {\doibase
  10.1002/adma.201806280} {\bibfield  {journal} {\bibinfo  {journal} {Adv.
  Mater.}\ }\textbf {\bibinfo {volume} {31}},\ \bibinfo {pages} {1806280}
  (\bibinfo {year} {2019})}\BibitemShut {NoStop}%
\bibitem [{\citenamefont {Greer}\ \emph {et~al.}(2013)\citenamefont {Greer},
  \citenamefont {Cheng},\ and\ \citenamefont {Ma}}]{GreerCheng2013}%
  \BibitemOpen
  \bibfield  {author} {\bibinfo {author} {\bibfnamefont {A.~L.}\ \bibnamefont
  {Greer}}, \bibinfo {author} {\bibfnamefont {Y.~Q.}\ \bibnamefont {Cheng}}, \
  and\ \bibinfo {author} {\bibfnamefont {E.}~\bibnamefont {Ma}},\ }\href
  {\doibase 10.1016/j.mser.2013.04.001} {\bibfield  {journal} {\bibinfo
  {journal} {Mater sci eng: R: Reports}\ }\textbf {\bibinfo {volume} {74}},\
  \bibinfo {pages} {71} (\bibinfo {year} {2013})}\BibitemShut {NoStop}%
\end{thebibliography}%

\end{document}